\documentclass[twoside,12pt,a4paper]{article}
\usepackage[top=25mm,bottom=25mm,left=25mm,right=25mm]{geometry}
\usepackage{amsmath,amssymb,color,graphics,graphicx,bm}
\usepackage{epstopdf}
\usepackage{graphicx,array}
\usepackage{dcolumn}
\usepackage{setspace}
\usepackage[utf8]{inputenc}
\usepackage{authblk}

\usepackage{scalerel}
\usepackage{hyperref}
\usepackage{relsize}
\usepackage{multicol}
\usepackage{blindtext, subfig}
\usepackage{float} 
\usepackage{natbib} 
\usepackage[toc]{appendix}
\numberwithin{equation}{section}

\pdfoutput=1
\title{Contact mechanics of adhesive beams. \\Part I: Moderate indentation.}
\date{}
\author{V. S. Punati\thanks{Electronic address: punati@iitk.ac.in; punati.iitk@gmail.com} , I. Sharma and P. Wahi}
\affil{Mechanics \& Applied Mathematics Group, Department of Mechanical Engineering, \\ Indian Institute of Technology Kanpur,  Kanpur, Uttar Pradesh, India - 208016.}

\begin{document}
\maketitle

\begin{abstract}
We investigate the contact of a rigid cylindrical punch with an adhesive beam mounted on flexible end supports. Adhesion is modeled through an adhesive zone model. The resulting Fredholm integral equation of the first kind is solved by a Galerkin projection method in terms of Chebyshev polynomials. Results are reported for several combinations of adhesive strengths, beam thickness, and support flexibilities characterized through torsional and vertical translational spring stiffnesses. Special attention is paid to the important extreme cases of clamped and simply supported beams. The popular Johnson-Kendall-Roberts (JKR) model for adhesion is obtained as a limit of the adhesive zone model. Finally, we compare our predictions with preliminary experiments and also demonstrate the utility of our approach in modeling complex structural adhesives.
\end{abstract}

\section{Introduction}
\begin{figure}[htbp]
\centering
\includegraphics[scale=1] {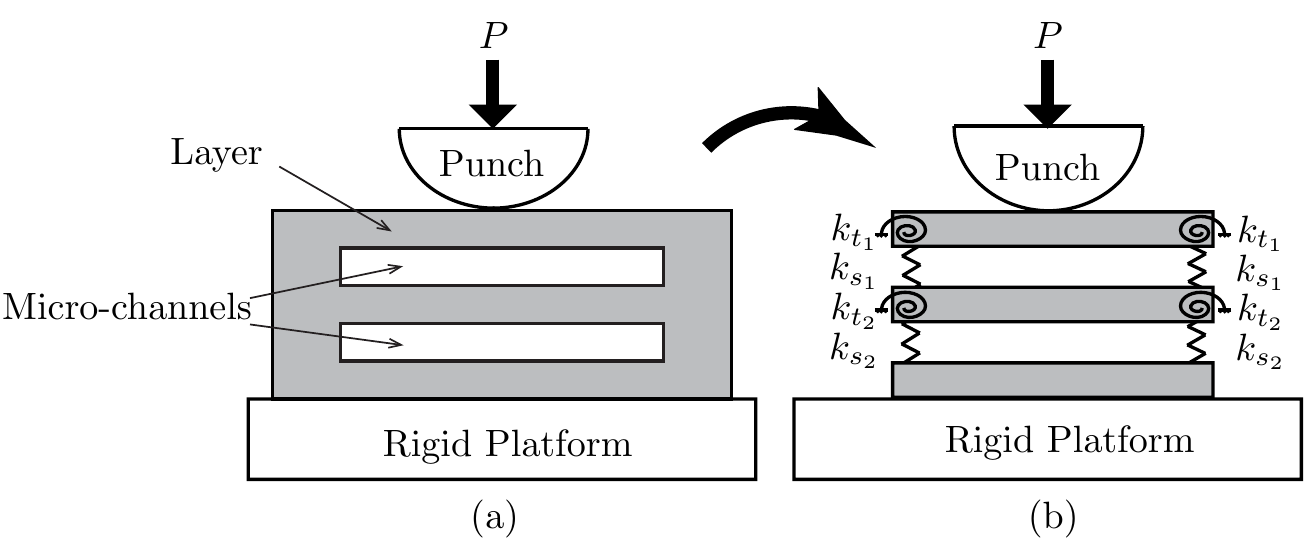}
\caption{(a) Structural adhesive designed by \cite{Arul2008bioinspired}. (b) Mechanical model of the structural adhesive in (a) employing an interconnected stack of adhesive beams. The rigidity of the vertical walls is modeled through torsional (stiffness $k_t$) and vertical translational (stiffness $k_s$) springs, as shown. The system is indented by a rigid punch, pressed down by the force $P$.}
\label{fig:struct_adhesive_model}
\end{figure}

\begin{figure}[htbp]
\centering
\includegraphics[scale=1] {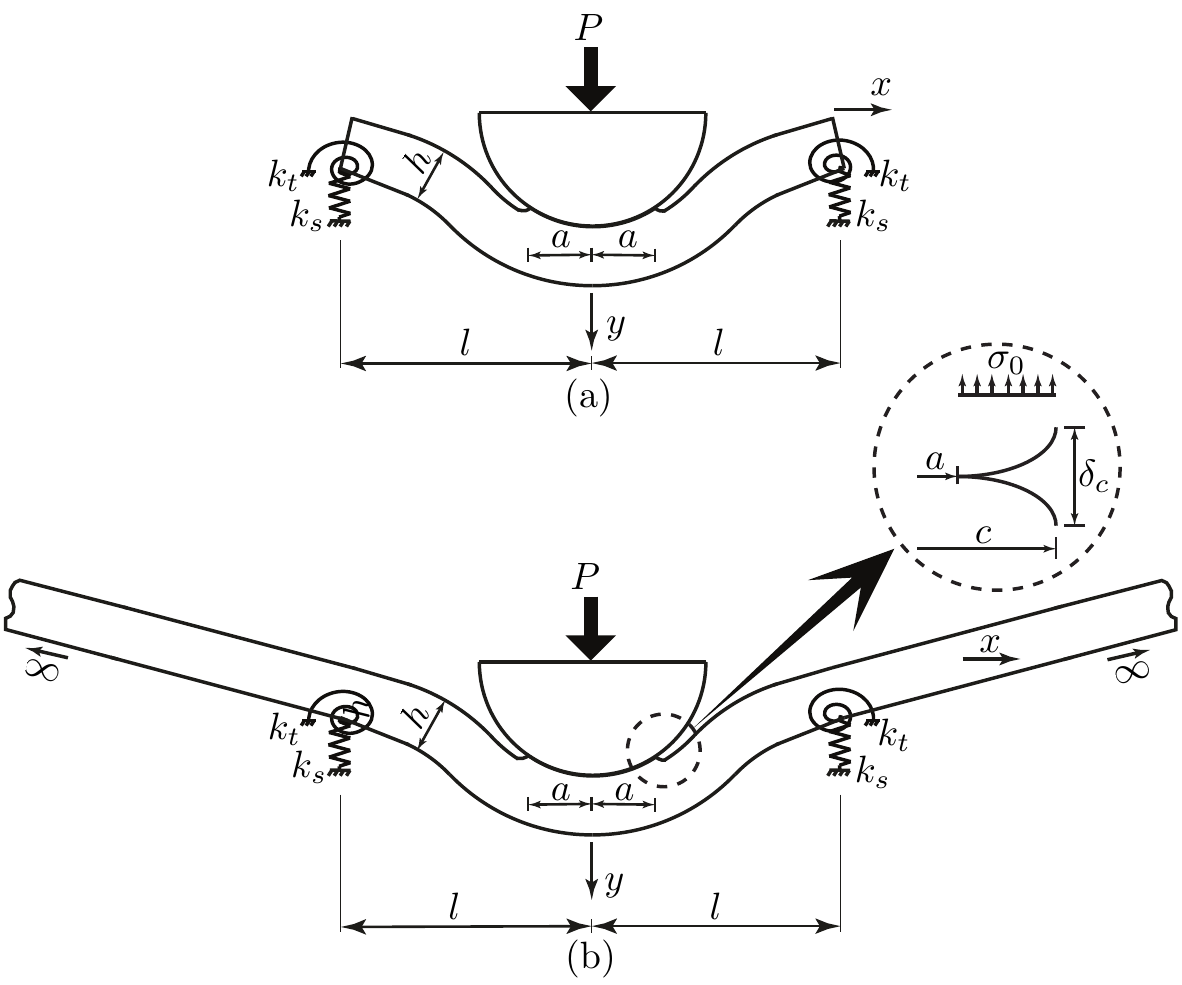}
\caption{(a) Indentation by a rigid cylindrical punch of an adhesive beam resting upon flexible supports. The flexible supports are modeled through torsional and vertical translational springs with stiffnesses $k_t$ and $k_s$, respectively. (b) Mathematical model of the indentation process shown in (a). The inset shows details of the adhesive zone active near the contact edges; see text for details. The vertical deflection is exaggerated for ease of representation.}
\label{fig:struct_adhesive_model_text2}
\end{figure}

Research in patterned adhesives is often motivated by the structures of natural adhesives, such as those present in the feet of gekkos; see e.g. \cite{Hiller1976}, and \cite{Arul2008bioinspired}. In conventional adhesives, such as thin, sticky tapes, only the top and bottom surfaces are active. However, multiple surfaces may be activated with appropriate patterning. With more surfaces participating in the adhesion process the adhesives show increased hysteresis and, so, better performance. One example of a patterned adhesive is the structural adhesive shown in Fig.~\ref{fig:struct_adhesive_model}(a), which was developed by \cite{Arul2008bioinspired}. Figure~\ref{fig:struct_adhesive_model}(b) shows a possible mechanical model of the structural adhesive of Fig.~\ref{fig:struct_adhesive_model}(a) that utilizes several interacting adhesive beams. This motivates the goal of this paper, which is to investigate the adhesive contact of a beam; see Fig.~\ref{fig:struct_adhesive_model_text2}(a).

Contact with a half-space has been well studied over the past century, and we refer the reader to  \cite{alexandrov2001three}, and also the texts of \cite{galin2008contact}, \cite{Gladwell1980contact}, \cite{Johnson1985contact} and \cite{Hills1993mechanics}. At the same time, the contact of thin layers is an active area of research in view of applications to electronics and computer industry; see, e.g. \cite{barthel2007adhesive}, and \cite{Dalmeya2012contact}. In contrast, the indentation of beams is much less studied.

Seeking the solution to the indentation of a beam through a strength-of-materials approach overlooks the local contact mechanics. To probe the latter, it is necessary to formulate an elasticity problem with appropriate boundary conditions. This is typically a complex problem, and has prompted some alternate approaches to adhesion\emph{less} contact, as discussed in the next paragraph. To the best of our knowledge there is \emph{no} available work on the adhesive contact of beams.

Two-dimensional adhesionless indentation of a beam has been studied in the past by \cite{keer1970bending}, \cite{Keer1983smooth},  \cite{keer1986smooth}, \cite{Sankar1983}, \cite{sun1985smooth}, and \cite{Kim2014}. \cite{Keer1983smooth} modeled the beam as a linear elastic layer of infinite extent with frictionless bottom and top surfaces. First, the elasticity problem was solved thorugh Fourier transforms, see e.g. \citet[p. 395-414]{Sneddon1995fourier}. Then, employing the Hankel transform (\citealt[p.~213]{Gladwell1980contact}) for the pressure distribution, and asymptotically matching the far-field displacements with those obtained from Euler-Bernoulli beam theory, a Fredholm integral equation of the second kind was obtained. This equation was solved numerically. 

\cite{Sankar1983} employed Fourier series in their investigation of adhesionless contact with beams of finite length. Their results were in good agreement with those of \cite{Keer1983smooth}. Recently, \cite{Kim2014} studied beam indentation through asymptotics. Finite element (FE) simulations were also carried out. The contact parameters, i.e. contact area and the total load acting on the punch, obtained through asymptotics, matched results of FE simulations well.  In all these studies, the interaction of the beam with the punch is non-adhesive. However, extending these methods to adhesive beams is difficult due to the presence of several iterated integral transforms. Here, we propose an alternative approach to the contact of a beam with a rigid punch, shown in Fig.~\ref{fig:struct_adhesive_model_text2}(a), that models both non-adhesive and adhesive interactions in a straightforward manner. 

Adhesion is modeled through the introduction of an adhesive zone that extends beyond the contact zone, as shown in the inset of Fig.~\ref{fig:struct_adhesive_model_text2}(b). An adhesive zone model allows us to investigate the effect of adhesion by admitting continuous variations in it's strength.  The popular JKR \citep{johnson1971surface} and DMT \citep{derjaguin1975effect} models for adhesion are obtained as special cases. The non-adhesive case (\textquoteleft Hertzian contact\textquoteright)  is found by setting the adhesion strength to zero. 

This paper is organized as follows. We start by formulating a mathematical model for the adhesive contact of the beam shown in Fig.~\ref{fig:struct_adhesive_model_text2}(a). We will obtain a Fredholm integral equation of the first kind that relates the contact pressure distribution with the vertical displacement in the contact patch. Relevant conditions at the edge of the contact zone will also be derived. This integral equation is then solved numerically employing Galerkin projections in terms of Chebyshev polynomials. Theoretical results are compared with FE simulations whenever possible. We then explore the effect of various parameters in the problem, viz. flexibility of end supports, strength of adhesion, the beam's geometry, etc. We close with a comparison with preliminary experimental results, and an application of our solution to complex structural adhesives, such as the one shown in Fig.~\ref{fig:struct_adhesive_model}(a). 

\section{Mathematical model}
\label{sec:Mathematical_model}

We begin by extending the beam of Fig.~\ref{fig:struct_adhesive_model_text2}(a) beyond the supports to infinity, as shown in Fig.~\ref{fig:struct_adhesive_model_text2}(b). The extension is done in a manner consistent with the kinematic and kinetic constraints imposed by the supports. Thus, the beam is extended linearly along its slope at the supports. The beam may now be thought of as a linear elastic layer of infinite length with thickness $h$. The beam is isotropic and homogeneous, with Young's modulus $E$  and Poisson's ratio $\nu$. The top and bottom surfaces of the beam are frictionless.

During indentation, a normal traction distribution $-P_c(x)$ acts on the top surface \footnote{The negative sign is introduced in order to report compressive pressure during contact as positive.}. At this time, the vertical displacement of the bottom surface is $v_b(x)$, which is typically \textit{not} known. When the contact area is less than the beam's thickness, we assume that $v_b(x)$ may be approximated by the displacement obtained from Euler-Bernoulli beam theory when a point load of magnitude $P$ acts at the center of the top surface, as depicted in Fig.~\ref{fig:fixedbeam_ptload_disp}. The details of how $v_b(x)$ is calculated for a beam on flexible supports are provided in App.~\ref{sec:Appendix_beamtheory}. 

The governing equations for the horizontal ($u$) and vertical ($v$) displacements in the extended beam (elastic layer) of Fig.~\ref{fig:struct_adhesive_model_text2}(b), assuming plane strain, are given by
\begin{subequations}
\label{Navier_equatons}
\begin{alignat}{1}
\frac{2 \left( 1- \nu \right) }{1-2 \nu} \frac{\partial^2 u}{\partial x^2} + \frac{\partial^2 u}{\partial y^2} + \frac{1}{1-2 \nu} \frac{\partial^2 v}{\partial x \partial y} &= 0 \label{Navier_eqn1}  \\
\text{and} \quad
\frac{\partial^2 v}{\partial x^2} +\frac{2 \left( 1- \nu \right) }{1-2 \nu} \frac{\partial^2 v}{\partial y^2} + \frac{1}{1-2 \nu} \frac{\partial^2 u}{\partial x \partial y} &= 0, \label{Navier_eqn2}
\end{alignat}
\end{subequations}
which reflect horizontal and vertical linear momentum balance, respectively; see, e.g. \citet[p.~241]{Timoshenko1970} or \citet[p.~125]{Sadd2010elasticity}.
The boundary conditions may be taken to be
\begin{subequations}
\label{BC_for_contact_x}
\begin{alignat}{3}
\sigma_{xy} &= 0,  &\quad \sigma_{yy} &= -P_c(x)  &\quad  \text{on the top surface, i.e. at $y=0$,} \label{BC_y0_x} \\
\text{and }\quad 
 \sigma_{xy} &= 0, &\quad v &= v_b(x)  &\quad \text{on the bottom surface, i.e. at $y=h$}. \label{BC_yh_x}
\end{alignat}
\end{subequations}
We now follow \citet[p.~402]{Sneddon1995fourier} to map the above problem into Fourier space by transforming the $x$ coordinate.  Solving for the  vertical displacement in Fourier space and, then, taking the inverse Fourier transform yields the following integral equation for the vertical displacement of the top surface:
\begin{align}
\label{v_x0}
v \left( x, 0 \right) =&  - \frac{2}{\pi E^*} \int\limits_{0}^{\infty} \bar{P}_c \left( \xi \right)\, \frac{ \sinh^2{\xi \, h}}{ \xi \left( \xi \, h + \sinh{\xi \, h } \cosh{\xi \, h} \right)} \, \cos{\xi x} \, d \xi \nonumber \\
& + \frac{1}{\pi} \int\limits_{0}^{\infty} \bar{v}_b \left( \xi \right) \, \frac {\sinh{ \xi \, h} + \xi\,h \cosh{ \xi\,h } }{ \xi \, h + \sinh{ \xi \, h} \cosh{ \xi \, h }} \, \cos{\xi x} \, d \xi,
\end{align}
where $E^* = E/ \left( 1 - \nu^2\right)$, 
\begin{equation}
\bar{P}_c \left( \xi \right) = \int\limits_{-\infty}^{\infty} - P_c \left( t \right) \cos{\xi t } \, dt \quad \text{and} \quad
\bar{v}_b \left( \xi \right)  = \int\limits_{-\infty}^{\infty} v_b \left( t \right) \cos{\xi t } \, dt .
\label{eqn:P_v_Fourier}
\end{equation}
Appendix \ref{sec:top_surface_disp} provides details of how \eqref{v_x0} is obtained.
For non-dimensionalizing it is more convenient to rewrite \eqref{v_x0} as
\begin{align}
v \left( x, 0 \right) =& \frac{2}{\pi E^*} \int\limits_{0}^{\infty} \int\limits_{-\infty}^{\infty} P_c \left( t \right) \cos{\xi t } \, dt \, \,  \frac{ \sinh^2{\xi \, h}}{ \xi \left( \xi \, h + \sinh{\xi \, h } \cosh{\xi \, h} \right) } \, \cos{\xi x} \, d \xi \nonumber \\
& + \frac{1}{\pi} \int\limits_{0}^{\infty} \int\limits_{-\infty}^{\infty} v_b \left( t \right) \cos{\xi t } \, dt \, \frac {\sinh{ \xi \, h} + \xi\,h \cosh{ \xi\,h } }{ \xi \, h + \sinh{ \xi \, h} \cosh{ \xi \, h }} \, \cos{\xi x} \, d \xi,
\label{eqn:v_x_0}
\end{align}
where we have invoked definitions \eqref{eqn:P_v_Fourier} of $\bar{P}_c$ and $\bar{v}_b$.

In contact problems the vertical displacement within the contact region is constrained. For example, during indentation with a rigid punch, the surface in the contact region must conform to the shape of the punch. We now approximate the profile of the cylindrical punch of radius $R$ as a parabola in the contact region, as is appropriate if the indentation depth, and the dimensions of the contact region are much smaller than the radius of curvature of the punch. We set $\delta$ to be the vertical displacement of the punch. This allows us to write the vertical displacement of the beam's top surface within the contact region as
\begin{equation}
\label{eqn:contactzone_disp_x}
v(x,0) = \delta-\frac{x^2}{2R}, \quad  \, -a \leq x \leq a,
\end{equation}
where the contact region lies between $-a$ and $a$.

During contact, the pressure on the beam's top surface depends also on the adhesive interaction between the beam and the punch. This adhesive interaction is, in turn, introduced through the presence of an adhesive zone; see inset of Fig.~\ref{fig:struct_adhesive_model_text2}(b).  Within the adhesive zone the adhesive interaction is modeled through a Dugdale-Barenblatt model \citep{Maugis1992adhesion}, which assumes the adhesion to be of constant strength $\sigma_0$. Thus, we may write the force distribution on the beam's top surface as
\begin{equation}
\label{eqn:Pfn}
P_c \left( x \right) = \left\{
\begin{array}{ll}
p \left( x \right),  &\quad |x|  \leq a \\
-\sigma_0, & \quad a < |x| \leq c \\
0, & \quad |x| > c,
\end{array}
\right.
\end{equation}
where $c$ locates the outer edge of the adhesive zone; see inset in Fig.~\ref{fig:struct_adhesive_model_text2}(b). Adhesive zones were introduced by \cite{Maugis1992adhesion} in order to avoid the singularity in the pressure at the contact edge ($x=\pm a$) found in JKR theory. For this, it is also necessary that there be no discontinuity in the contact pressure at the contact edge, i.e.
\begin{equation}
\label{eqn:contact_end_pressure}
\lim_{ x  \rightarrow \pm a^{-}}  p \left( x \right) = -\sigma_0.
\end{equation}
To close our mathematical description we require an additional equation to compute the extent $c$ of the adhesive zone. This is obtained by equating the energy release rate computed from the $J-$integral \citep{rice1968path} and the work of adhesion $w$, which leads to
\begin{equation}
\label{eqn:Griffith_eqn}
\sigma_0 \delta_c = w,
\end{equation}
where 
\begin{equation}
\label{eqn:air_gap}
\delta_c = \left( c^2/2R \right) - \delta + v_c
\end{equation} 
is the air-gap at which the adhesive forces vanish and $v _c = v\left( c,0 \right)$;  see inset in Fig.~\ref{fig:struct_adhesive_model_text2}(b). 

During non-adhesive indentation \eqref{eqn:air_gap} is automatically satisfied as $\sigma_0 = 0 =w$. The JKR approximation is obtained in the limits $\sigma_0 \rightarrow \infty$ and $c \rightarrow a$, at which the energy balance \eqref{eqn:Griffith_eqn} becomes
\begin{equation}
\label{fracture_Griffith}
\frac{K_1^2}{2E^*} = w,
\end{equation}
where
\begin{equation}
\label{stress_intensity factor}
K_1=-\lim_{x\rightarrow a^{-}} \sqrt{2 \pi \left( a-x \right)}  p \left( x \right)
\end{equation}
is the \emph{stress intensity factor} that measures the strength of the square root singularity in the pressure at the contact edge; see \cite{Maugis1992adhesion}. This is equivalent to Griffith's criterion in fracture mechanics; see e.g. \citet[p.~168]{kanninen1985advanced}. In this limit, we do not require the contact pressure end condition \eqref{eqn:contact_end_pressure}.

Finally, the total load acting on the punch is found by integrating the normal traction over the top surface of the beam:
\begin{equation}
\label{eqn:total_load}
P = \int\limits_{-\infty}^{\infty} P_c \left( x \right) \, dx = \int\limits_{-a}^{a} p \left( x \right) dx - 2 \, \sigma_0  \left( c - a \right).
\end{equation}

\section{Non-dimensionalization}
\label{sec:Non-dimensionalization}
We introduce the following non-dimensional parameters: 
\begin{gather*}
A = \frac{a}{l} ;  \quad  \varphi \left( \tau \right) = \frac{aRl}{K h^3} p \left( a \bar{\tau} \right) ; \quad  \bar{P} = \frac{P Rl}{K h^3};   \quad k_t^f = \frac{k_t l } {E I};  \quad k_s^f =\frac{ k_s l^3 }{E I}; \\
\Delta =  \frac{\delta R}{l^2}; \quad L= \frac{l}{R}; \quad \lambda = 2 \sigma_0 \left( \frac{R}{\pi w K^2} \right)^{1/3};  \quad  m = \left( \frac{\pi w}{RK} \right)^{1/3}, 
\end{gather*}
where $K=4 \, E^*/3$ and $I = h^3/12$ is the beam's area moment of inertia. We also define the scaled variables
\begin{gather*}
\left\{ \bar{x} , \bar{\tau}, \bar{c}, \bar{\gamma}  \right\} = \frac{1}{a} \, \left\{ x, t, c, h \right\}; \quad
  \left\{ \hat{\tau} , \hat{\gamma}  \right\} =  \frac{1}{l} \, \left\{  t,  h \right\}; \quad 
\left\{ \omega, \bar{\omega}, \hat{\omega}  \right\} = \left\{ \xi h, \frac{\omega}{\bar{\gamma}},  \frac{\omega}{\hat{\gamma}}\right\};   \\ 
\left\{ \vartheta \left( \bar{x}, 0 \right), \vartheta_b \left( \hat{\tau} \right) \right\} = \frac{R}{l^2} \left\{ v \left( \bar{x}, 0 \right), v_b \left( l \hat{\tau} \right) \right\} ; \quad
 \bar{I} = \frac{I}{h^3} = \frac{1}{12}; \quad
 \Phi \left( \tau \right) = \frac{a R l}{K h^3}  \, P_c \left( a \bar{\tau} \right).
\end{gather*}
In terms of these variables the non-dimensional vertical displacement of the top surface \eqref{eqn:v_x_0} becomes
\begin{align}
\vartheta \left( \bar{x}, 0 \right) =& \frac{8 \hat{\gamma}^3}{3 \pi} \int\limits_{0}^{\infty} \int\limits_{-\infty}^{\infty} \Phi \left( \bar{\tau} \right) \cos \left( \bar{\omega} \bar{\tau} \right) \, d\bar{\tau} \, \, K_1 \left( \bar{\omega}, \bar{x} \right) \,  \, d \omega  \quad + \nonumber \\
& \frac{1}{\pi \hat{\gamma}} \int\limits_{0}^{\infty} \int\limits_{-\infty}^{\infty} \vartheta_b \left( \hat{\tau} \right) \cos{ \hat{\omega} \hat{\tau} } \, d \hat{\tau} \, \,  K_2 \left( \bar{\omega}, \bar{x} \right)  \, d \omega,
\label{the_final_2}
\end{align} 
with the kernels 
\begin{eqnarray}
 K_1 \left( \bar{\omega}, \bar{x} \right) &=& \frac{ \sinh^2{\omega}}{ \omega \left( \omega + \sinh{\omega } \cosh{\omega} \right) } \, \cos \left( \bar{\omega} \bar{x}  \right) \nonumber \\ \text{and} \hspace{5cm}
 K_2 \left( \bar{\omega}, \bar{x} \right) &=& \frac{\sinh{ \omega} + \omega \cosh{ \omega } }{ \omega + \sinh{ \omega} \cosh{ \omega }} \, \cos \left( \bar{\omega} \bar{x}  \right).  \hspace{5cm} \nonumber
\end{eqnarray}
From \eqref{eqn:Pfn}, we obtain the non-dimensional pressure on the beam's top surface:
\begin{equation}
\label{eqn:Pfn_nondim} 
\Phi \left( \bar{\tau} \right) = \left\{
\begin{array}{ll}
\varphi \left( \bar{\tau} \right), & \quad |\bar{\tau}| \leq 1 \\
-\lambda A m / 2 \hat{\gamma}^3 L,  & \quad  1 < |\bar{\tau}| \leq \bar{c} \\
0, & \quad  |\bar{\tau}| > \bar{c}.
\end{array}
\right.
\end{equation}
Combining \eqref{the_final_2} and \eqref{eqn:Pfn_nondim} yields
\begin{align}
\vartheta \left( \bar{x}, 0 \right)   = & - \frac{8 \hat{\gamma}^3}{3 \pi} \int\limits_{0}^{\infty} \bar{\varphi} \left( \bar{\omega}  \right) \, K_1 \left( \bar{\omega},\bar{x}  \right)
 \,  \, d \omega  - \frac{8 \lambda A m}{3 \pi L}   \int\limits_{0}^{\infty}  \bar{\varphi}_0 \left( \bar{\omega}  \right) \,  K_1 \left(\bar{\omega},\bar{x}  \right) \, \, d \omega  \nonumber \\
 &  + \frac{1}{\pi \hat{\gamma}} \int\limits_{0}^{\infty} \hat{\vartheta}_b \left( \hat{\omega} \right) \,  K_2 \left( \bar{\omega},\bar{x}  \right) \, \, d \omega,
 \label{Int_eqn}
\end{align}  
with
\begin{equation}
\bar{\varphi} \left( \bar{\omega}  \right) = - \int\limits_{-1}^{1} \varphi \left( \bar{\tau} \right) \cos \left( \bar{\omega}  \, \bar{\tau} \right) \, d\bar{\tau}, \quad
\bar{\varphi}_0 \left( \bar{\omega}  \right) = \int\limits_{1}^{\bar{c}}\cos \left( \bar{\omega}  \, \bar{\tau} \right) \, d\bar{\tau}  \quad \text{and } \,
\hat{\vartheta}_b \left( \hat{\omega} \right) = \int\limits_{-\infty}^{\infty} \vartheta_b \left( \hat{\tau} \right) \cos \left( \hat{\omega}   \hat{\tau} \right) \, d \hat{\tau}. \label{integrals_in_omega}
\end{equation}
Non-dimensionalizing the displacement in the contact region \eqref{eqn:contactzone_disp_x}, the contact pressure end condition \eqref{eqn:contact_end_pressure}, and the energy equation \eqref{eqn:Griffith_eqn} lead to, respectively,
\begin{flalign}
&& \vartheta (\bar{x},0) &= \Delta  - \frac{1}{2} \, \bar{x}^2 A^2 \quad \text{for } \quad -1 \leq \bar{x} \leq 1, \label{contactzone_disp_nondim} &&\\
&& \varphi \left( \pm 1 \right) &= -\frac{\lambda \, A m}{2 \, \hat{\gamma}^3 L} \label{contact_end_pressure_nondim} && \\
\text{and } &&
1 &= \frac{\pi \lambda L^2}{2m^2} \left[ \frac{\bar{c}^2 A^2}{2}  - \Delta  + \vartheta_c  \right], \label{Griffith_Eqn_nondim} &&
\end{flalign}
where $\vartheta_c = \vartheta \left( \bar{c},0 \right)$ and $\Delta$ is the non-dimensional displacement of the punch. Combining \eqref{fracture_Griffith} and \eqref{stress_intensity factor}, and non-dimensionalizing, we obtain
\begin{equation}
\label{fracture_Griffith_nondim}
\lim_{\bar{x} \rightarrow 1^{-}} \sqrt{\left( 1 - \bar{x} \right)}  \varphi \left(  \bar{x} \right) = -\frac{m}{2 \pi L} \left( \frac{l}{h} \right)^3 \sqrt{\frac{3 A m}{L}}, 
\end{equation}
which replaces \eqref{contact_end_pressure_nondim} and \eqref{Griffith_Eqn_nondim} whenever we invoke the JKR approximation.

 The total non-dimensional load acting on the punch is found from \eqref{eqn:total_load}:
\begin{equation}
\label{load_nondim}
\bar{P} = \int\limits_{-1}^{1} \varphi \left( \bar{\tau} \right) \, d \bar{\tau} - \frac{\lambda A m}{\hat{\gamma}^3 L} \left( \bar{c} - 1 \right).
\end{equation}
Finally,  evaluating \eqref{Int_eqn} in the contact region, i.e. for $-1 \le \bar{x} \le 1$, and employing \eqref{contactzone_disp_nondim} we obtain
\begin{align}
\label{Int_eqn_final}
\Delta  - \frac{1}{2} \, \bar{x}^2 A^2  = & - \frac{8 \hat{\gamma}^3}{3 \pi} \int\limits_{0}^{\infty} \bar{\varphi} \left( \bar{\omega}  \right) \, K_1 \left( \bar{\omega},\bar{x}  \right)
 \,  \, d \omega  - \frac{8 \lambda A m}{3 \pi L}   \int\limits_{0}^{\infty}  \bar{\varphi}_0 \left( \bar{\omega}  \right) \,  K_1 \left(\bar{\omega},\bar{x}  \right) \, \, d \omega  \nonumber \\
 &  + \frac{1}{\pi \hat{\gamma}} \int\limits_{0}^{\infty} \hat{\vartheta}_b \left( \hat{\omega} \right) \,  K_2 \left( \bar{\omega},\bar{x}  \right) \, \, d \omega.
\end{align} 
This is a Fredholm integral equation of first kind; see \citet[p.~573]{polyanin2008handbook}. We now solve the above equation, along with boundary conditions \eqref{contact_end_pressure_nondim} and \eqref{Griffith_Eqn_nondim}, for the contact pressure $\varphi \left( \bar{x} \right)$, displacement $\Delta$ and the location $\bar{c}$ of the adhesive zone's edge, at a given contact area $A$. 

\section{Numerical solution}
\label{sec:Numerical_solution}
The integral equation \eqref{Int_eqn_final}  does not admit an analytical solution due to the complex kernels present. Thus, we solve it numerically. To this end, we approximate the unknown pressure distribution $p \left( x \right)$ in the contact region through a series of Chebyshev polynomials. Chebyshev polynomials are chosen due to their spectral convergence; see \citet[p.~63]{mason2003book}.

The unknown non-dimensional pressure distribution is expressed as a series of Chebyshev polynomials of the first kind, viz. 
\begin{equation}
\label{phi_cheby}
\varphi \left( \bar{\tau} \right) = - \frac{ \lambda A m}{2\hat{\gamma}^3 L} + \frac{1}{\sqrt{1-\bar{\tau}^2}} \sum_{n=0}^{N} b_{2n} T_{2n} \left( \bar{\tau} \right),
\end{equation}
where $b_{2n}$ are unknown constants that are to be found. 
Only even Chebyshev polynomials are considered as the problem is symmetric about the origin. The constant term in the approximation is introduced to explicitly account for the continuity condition \eqref{contact_end_pressure_nondim} that is imposed on the contact pressure at the edge of the contact zone.  

Employing \eqref{phi_cheby} to evaluate the integrals $\bar{\varphi} \left( \bar{\omega} \right)$ and $\bar{\varphi}_0 \left( \bar{\omega}  \right)$ from \eqref {integrals_in_omega}, we obtain, respectively,  
\begin{flalign}
&& \bar{\varphi} \left( \bar{\omega}  \right) &= \frac{ \lambda A m}{\hat{\gamma}^3 L} \, \frac{\sin{\bar{\omega} } }{\bar{\omega}} - \sum_{n=0}^{N} b_{2n} \alpha_{2n} \left( \bar{\omega} \right)  \label{phi_cheby_omega} &&\\
\text{and } &&  
\bar{\varphi}_0 \left( \bar{\omega}  \right) &= \frac{1}{\bar{\omega}} \, \left( -\sin{ \bar{\omega} } +\sin{\bar{\omega} \bar{c} } \right), && 
 \label{phi_tilde_calc}
\end{flalign}
where
\begin{equation}
\label{alpha_cheby_2}
\alpha_{2n} \left( \bar{\omega} \right) = \int\limits_{-1}^{1} \frac{1}{\sqrt{ \left( 1-\tau^2 \right)}} T_{2n} \left( \bar{\tau} \right) \cos{ \bar{\omega} \bar{\tau} } \: \: d\bar{\tau}.
\end{equation}
Appendix \ref{sec:Appendix_alphan} provides details of how $\alpha_{2n} \left( \bar{\omega} \right)$ are computed.
Combining  \eqref{load_nondim} and \eqref{phi_cheby}, we find the total load acting on the punch to be
\begin{equation}
\label{total_load_cheby}
\bar{P} = \pi b_0 - \frac{ \lambda A m}{\hat{\gamma}^3 L} \bar{c}.
\end{equation}
The displacement of the beam's bottom surface $\bar{\vartheta}_b \left( \hat{\omega} \right) $ may, with \eqref{total_load_cheby}, be written as
 \begin{align}
\frac{1}{\pi \hat{\gamma}} \hat{\vartheta}_b \left( \hat{\omega} \right)  = \frac{4}{3  \hat{\gamma} \bar{I} \left( 1-\nu^2 \right) }  \left( b_0  - \frac{\lambda A m \bar{c} }{\pi \hat{\gamma}^3  L} \right) \hat{\vartheta}_p \left( \hat{\omega}  \right), \label{V_omega}
\end{align}
where $\bar{I}$ is the scaled area moment of inertia, and the exact form of $\bar{\vartheta}_p$ depends upon how the beam is supported at its ends; see App.~\ref{sec:Appendix_beamtheory}.

The vertical displacement in the contact region may be expressed in terms of Chebyshev polynomials as
\begin{equation}
\label{Delta_tau}
\Delta  - \frac{1}{2} \, \bar{x}^2 A^2 = \sum_{n=0}^{N} a_{2n} T_{2n} \left( \bar{x} \right) = \left( \Delta - \frac{{A}^{2}}{4} \right) T_0 \left( \bar{x} \right) - \frac{A^2}{4} T_2 \left( \bar{x} \right).
\end{equation}
Employing expansions \eqref{phi_cheby}--\eqref{Delta_tau} in the integral equation \eqref{Int_eqn_final},  we obtain
\begin{align}
\sum_{n=0}^{N} a_{2n} T_{2n} \left( \bar{x} \right)  =& \frac{8 \hat{\gamma}^3}{3 \pi} \sum_{n=0}^{N} b_{2n} \mathcal{J}^{(1)}_{2n} \left( \bar{x}  \right) - \frac{8 \lambda A m}{3 \pi L}   \mathcal{J}^{(1)} \left( \bar{x}  \right) + 
\frac{4 }{3 \hat{\gamma} \bar{I} \left( 1-\nu^2 \right)}  \left( b_0  - \frac{\lambda A m \bar{c} }{ \pi \hat{\gamma}^3 L } \right) \mathcal{J}^{(2)} \left( \bar{x}  \right),
\label{Int_eqn_m4}
\end{align}
where
\begin{eqnarray}
\mathcal{J}^{(1)}_{2n} \left( \bar{x}  \right) &=& \int\limits_{0}^{\infty} \alpha_{2n} \left( \bar{\omega} \right) \, K_1 \left( \bar{\omega},\bar{x}  \right) \,  \, d \omega, \nonumber \\
\mathcal{J}^{(1)} \left( \bar{x}  \right) &=& \int\limits_{0}^{\infty} \frac{\sin{ \bar{\omega} \, \bar{c} } }{\bar{\omega}} \,   K_1 \left( \bar{\omega},\bar{x}  \right) \, \, d \omega  \nonumber \\ \text{and } \hspace{5cm}
\mathcal{J}^{(2)} \left( \bar{x}  \right) &=& \int\limits_{0}^{\infty} \hat{\vartheta}_p \left( \hat{\omega}\right) \,  K_2 \left( \bar{\omega},\bar{x}  \right) \, \, d \omega. \hspace{5cm} \nonumber
\end{eqnarray}
We evaluate the above integrals at any $\bar{x}$ through the Clenshaw-Curtis quadrature (\citealt[p.~196]{press1992numerical}). We now follow \citet[p.~267]{Gladwell1980contact} and utilize Galerkin projections to solve \eqref{Int_eqn_m4} for the unknown coefficients $b_{2n}$. To this end, we multiply both sides of \eqref{Int_eqn_m4} by $T_{2m} \left( \bar{x} \right) / \sqrt{1-\bar{x}^2} $, for $m=0,\cdots, N$, and integrate from $\bar{x}=-1$ to $\bar{x} =1$. This yields the following system of $N+1$ linear algebraic equations:
\begin{align}
\sum_{n=0}^{N} a_{2n} \mathcal{J}_{nm} = & \frac{8 \hat{\gamma}^3}{3 \pi} \sum_{n=0}^{N} b_{2n} \mathcal{J}^{(1)}_{nm} - 
\frac{8 \lambda A m}{3 \pi L}  \mathcal{J}^{(1)}_{m} +
 \frac{4}{3 \hat{\gamma} \bar{I} \left( 1-\nu^2 \right)} \left( b_0  - \frac{\lambda A m \bar{c} }{ \pi \hat{\gamma}^3 L} \right) 
\mathcal{J}^{(2)}_{m}, \label{Int_eqn_numerical_final}
\end{align}
where
\begin{alignat}{3}
 \mathcal{J}_{nm}  &= \int\limits_{-1}^{1} \frac{T_{2n} \left( \bar{x} \right) \,  T_{2m} \left( \bar{x} \right)}{\sqrt{1-\bar{x}^2}}  \, d \bar{x},  & \quad 
 \mathcal{J}^{(1)}_{nm}   &= \int\limits_{-1}^{1} \frac{\mathcal{J}^{(1)}_{2n} \left( \bar{x} \right) \, T_{2m} \left( \bar{x} \right)}{\sqrt{1-\bar{x}^2}} \, d \bar{x}, \nonumber \\
  \mathcal{J}^{(1)}_{m} &=  \int\limits_{-1}^{1} \frac{\mathcal{J}^{(1)} \left( \bar{x}  \right) \, T_{2m} \left( \bar{x} \right)}{\sqrt{1-\bar{x}^2}}  \,  d \bar{x} \nonumber  & \quad  \text{and} \hspace{0.25cm}
 \mathcal{J}^{(2)}_{m} &= \int\limits_{-1}^{1} \frac{\mathcal{J}^{(2)} \left( \bar{x}  \right) \, T_{2m} \left( \bar{x} \right)}{\sqrt{1-\bar{x}^2}} \, \, d \bar{x}. \nonumber
\label{Int_eqn_TT_final} 
\end{alignat}
The foregoing integrals are evaluated through a Gauss-Chebyshev quadrature (\citealt[p.~260]{Gladwell1980contact}).

Equations for $ b_{2n}$, $\Delta$ and $\bar{c}$ are now obtained. Employing the expansion \eqref{phi_cheby} in the contact pressure end condition \eqref{contact_end_pressure_nondim} yields 
\begin{equation}
\label{end_pressure}
b_{0} + b_{2} + \cdots + b_{2N} = 0.
\end{equation}
The energy balance \eqref{Griffith_Eqn_nondim} provides
\begin{equation}
\label{Griffith_Eqn_numerical}
\frac{\pi \lambda L^2}{2m^2} \left( \frac{\bar{c}^2 A^2}{2}  - \Delta  + \vartheta_c  \right) =1,
\end{equation}
where the non-dimensional displacement of the punch
\begin{align}
\Delta  =& \frac{8 \hat{\gamma}^3}{3 \pi} \sum_{n=0}^{N} b_{2n} \mathcal{J}^{(1)}_{2n} \left( 0  \right) - \frac{8 \lambda A m}{3 \pi L}   \mathcal{J}^{(1)} \left( 0  \right) +  
 \frac{4}{3 \hat{\gamma} \bar{I} \left( 1-\nu^2 \right)} \left( b_0 - \frac{\lambda A m \bar{c} }{\pi \hat{\gamma}^3 L} \right)   \mathcal{J}^{(2)} \left( 0  \right),
\label{Delta_eqn}
\end{align}
and the air gap at the end of the adhesive zone is
\begin{align}
\vartheta \left( \bar{c} \right) = &
\frac{8\hat{\gamma}^3}{3 \pi} \sum_{n=0}^{N} b_{2n} \mathcal{J}^{(1)}_{2n} \left( \bar{c}  \right) - \frac{8 \lambda A m}{3 \pi L}   \mathcal{J}^{(1)} \left( \bar{c}  \right) + 
 \frac{4}{3 \hat{\gamma} \bar{I} \left( 1-\nu^2 \right)}  \left( b_0 - \frac{\lambda A m \bar{c} }{\pi \hat{\gamma}^3 L} \right)   \mathcal{J}^{(2)} \left( \bar{c}  \right).
\label{V_eqn}
\end{align}

Finally, we have $N+1$ equations from \eqref{Int_eqn_numerical_final} and one each from \eqref{end_pressure} and \eqref{Griffith_Eqn_numerical}, for a total of $N+3$ equations. For a given contact area $A$, the total number of unknowns in this problem are also $N+3$: the unknown coefficients $b_{2n}$, with $n=0,\cdots, N$ in the expansion \eqref{phi_cheby} of the contact pressure $\varphi \left( \bar{x} \right)$, the displacement $\Delta$ of the punch, and the location $\bar{c}$ of the adhesive zone's edge. 

The system of equations \eqref{Int_eqn_numerical_final}--\eqref{Griffith_Eqn_numerical} are linear in $b_{2n}$ and $\Delta$, but \textit{non}-linear in $\bar{c}$; cf.~\eqref{Griffith_Eqn_numerical}. Thus, an iterative procedure is followed beginning with an initial guess for $\bar{c}$. At any $\bar{c}$, \eqref{Int_eqn_numerical_final} and \eqref{end_pressure}  are solved for $b_{2n}$ and $\Delta$. These $b_{2n}$ and $\Delta$ must satisfy \eqref{Griffith_Eqn_numerical} at the current value of $\bar{c}$. If \eqref{Griffith_Eqn_numerical} is not satisfied, then the value of $\bar{c}$ is updated through a Newton-Raphson root finding algorithm; see, e.g. \cite{Anindya}. We continue to iterate until a consistent set of $b_{2n}$, $\Delta$ and $\bar{c}$ is found. Care should be taken while finding the adhesive zone size $\bar{c}$, as it depends sensitively on the initial guess, and on the root finding algorithm that is employed. Once $b_{2n}$, $\Delta$ and $\bar{c}$ are found, the contact pressure $\varphi \left( \bar{x} \right)$ and the total load $\bar{P}$ may be obtained from \eqref{phi_cheby} and \eqref{total_load_cheby}, respectively.

\section{Finite element simulations}
\label{sec:FE_model}
For comparison later, we will also solve the non-adhesive (\textquoteleft Hertzian\textquoteright) contact of a rigid punch with a beam through the finite element (FE) method. We employ the commercial FE package ABAQUS. While ABAQUS does provide some cohesive zone models that may be employed to simulate adhesive contact, they are not easily compared with the Dugdale-Barenblatt model that we employ. Thus, we restrict comparisons with FE results to non-adhesive contact. We also limit FE simulations to clamped and simply supported beams.

In our FE simulations, the beam is modeled as a linear elastic layer with Young's modulus $E=2000$ MPa and Poisson's ratio $\nu=0.3$. The beam's thickness and half-span are taken as $h=4$ mm and $l=40$ mm, respectively. We note that these material properties are not typical of beams employed in structural adhesives, for example by \cite{Arul2008bioinspired}. However, these properties are selected as their magnitudes allow us to easily distinguish the effect of external inputs to the punch. 

The cylindrical punch has radius $R=225$ mm and Young's modulus $E_p=2 \times 10^{6}$ MPa - a thousand times the Young's modulus of the beam. A high $E_p$ is chosen in order to approximate a rigid punch. 

In our FE analysis, plane-strain elements are considered for both the beam and the punch. The load is applied on the punch. The remaining contact parameters, i.e. contact pressure, contact area, and the displacement of the punch, are obtained after post-processing the computation's output. These parameters are now compared with the semi-analytical results of Sec.~\ref{sec:Numerical_solution}.

\section{Results: Non-adhesive (\textquoteleft Hertzian\textquoteright) contact}
\label{sec: Comparision_with_ABAQUS}
We first investigate the adhesionless contact of a rigid punch with elastic beams. We will consider beams that are clamped, simply supported, or rest on flexible supports. Some results will be compared with FE simulations of Sec.~\ref{sec:FE_model}. 

Equations for the indentation of a non-adhesive beam are obtained by setting $\lambda=0$ in \eqref{Int_eqn_numerical_final}:
\begin{equation}
\sum_{n=0}^{N} a_{2n} \mathcal{J}_{nm} = \frac{8 \hat{\gamma}^3}{3 \pi} \sum_{n=0}^{N} b_{2n} \mathcal{J}^{ \left( 1 \right)}_{nm} + \frac{4 b_0 }{3 \hat{\gamma} \bar{I} \left( 1-\nu^2 \right)} \mathcal{J}^{ \left( 2 \right) }_{m} 
\quad \text{for } m=0,\cdots, N.
\label{Int_eqn_Hertz}
\end{equation}
The contact pressure vanishes at the edge $\bar{x} = \pm 1$ of the contact zone, so that \eqref{end_pressure} holds. The energy condition \eqref{Griffith_Eqn_numerical} is now redundant. The $N+2$ equations that comprise \eqref{end_pressure} and \eqref{Int_eqn_Hertz} are to be solved for the $N+2$ unknowns $b_{2n}$ $(n=0,\cdots,N)$ and $\Delta$ for a given choice of $A$. The contact pressure distribution and the total load are then found from \eqref{phi_cheby} and \eqref{total_load_cheby}, respectively, after setting $\lambda=0$.

Computations are carried out with $N=5$, i.e. the expansion \eqref{phi_cheby} is truncated at the Chebyshev polynomial $T_{10}$. 

\subsection{Clamped beam}
\label{Subsec: Non-adhesive fixed beam with ABAQUS}
A clamped beam is obtained in the limit of $k_s^f, \,  k_t^f \rightarrow \infty$. Thus, $\bar{\vartheta}_p \left( \hat{\omega} \right)$ is given by \eqref{four_defn_vp}, which then enters into the computation of $\mathcal{J}^{ \left( 2 \right) }_{m}$ in \eqref{Int_eqn_Hertz}. The unknown contact pressure distribution $\varphi \left( \bar{x} \right)$ is obtained by solving \eqref{Int_eqn_Hertz} and \eqref{end_pressure}. 

\begin{figure}[htbp]
\begin{center}
\subfloat[][]{\includegraphics[width=0.5\linewidth]{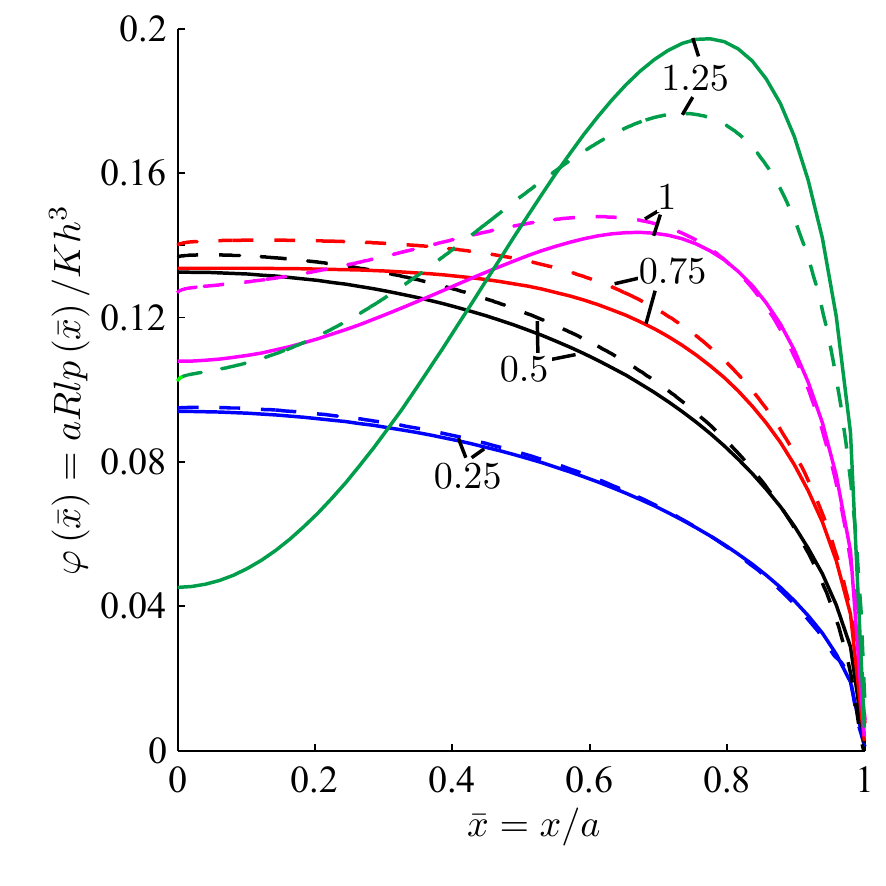}}
\subfloat[][]{\includegraphics[width=0.5\linewidth]{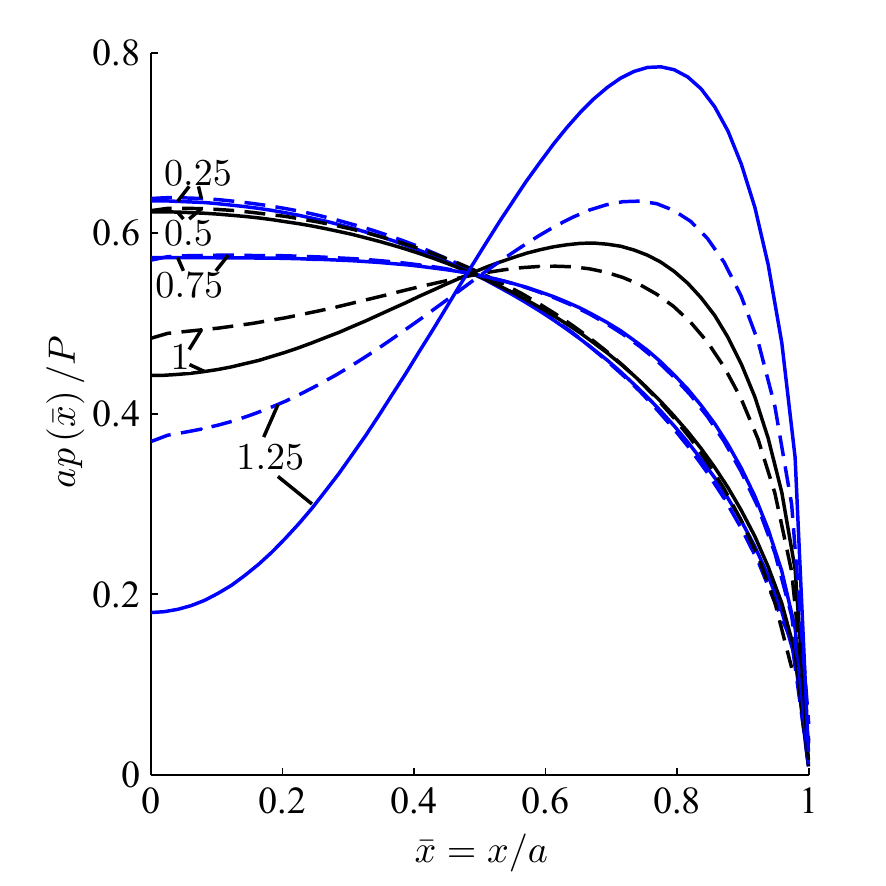}}
\caption{Non-adhesive contact of a clamped beam. Non-dimensional contact pressures (a) $ \varphi \left( \bar{x} \right) $ and (b) $a p \left( \bar{x} \right) / P$ are shown. The beam's slenderness ratio $l/h=1/10$. Several contact areas $a$ are investigated by varying $a/h$, which are noted next to their associated curves. Solid lines are results obtained from the semi-analytical procedure of Sec.~\ref{sec:Numerical_solution}. Dashed lines correspond to FE simulations of Sec.~\ref{sec:FE_model}.}
\label{clampedbeam_pressure_compare_FE}
\end{center}
\end{figure}

\begin{figure}[htbp]
\centering
\includegraphics[width=0.5\linewidth]{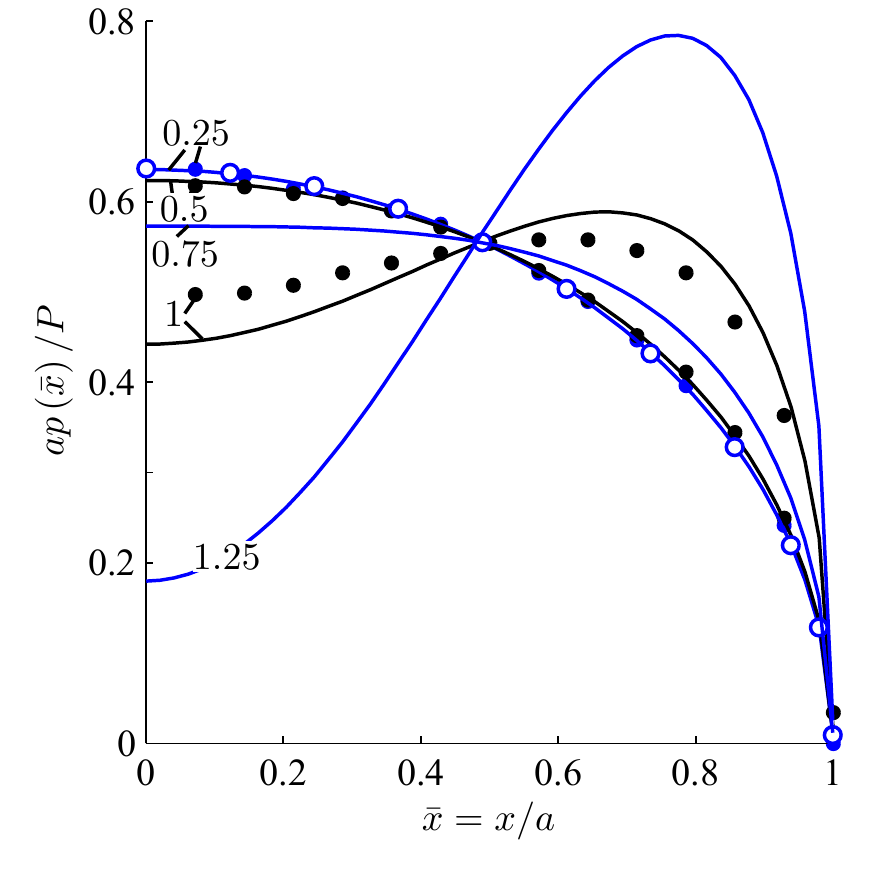}
\caption{Non-adhesive contact of a clamped beam. Non-dimensional contact pressure $a p \left( \bar{x} \right) / P$ is shown. The beam's slenderness ratio $l/h=10$. Several contact areas $a$ are investigated by varying $a/h$, which are noted next to their corresponding curves. Solid lines are results obtained from the semi-analytical procedure of Sec.~\ref{sec:Numerical_solution}. Open-circles represent the solution for an elastic half-space. Results of \cite{Keer1983smooth}, when available, are shown by filled circles.}
\label{clampedbeam_pressure_Halfspace_KM}
\end{figure}

We compare the results of our semi-analytical procedure of Sec.~\ref{sec:Numerical_solution} with FE simulations in Fig.~\ref{clampedbeam_pressure_compare_FE} and with the results of \cite{Keer1983smooth} in Fig.~\ref{clampedbeam_pressure_Halfspace_KM}.
We observe from Figs.~\ref{clampedbeam_pressure_compare_FE} and \ref{clampedbeam_pressure_Halfspace_KM} that, when the ratio $a/h$ of the contact area to the beam's thickness is low, the maximum contact pressure is obtained at the center of the contact region. We also find from Fig.~\ref{clampedbeam_pressure_Halfspace_KM} that at low $a/h$ ratios the pressure profiles are similar to the pressure distribution obtained for indentation into an elastic half-space.  Increasing the $a/h$ ratio -- which, for a given beam (fixed $h$ and $l$) corresponds to increasing the load, as the contact area increases -- causes the pressure at the center of the contact region to decrease, but increase near its ends; thus, the pressure profiles acquire a double-humped character. Our semi-analytical results are in good agreement with those of FE simulations, and the theoretical results of \cite{Keer1983smooth} for $a/h \lessapprox 1$; see Figs.~\ref{clampedbeam_pressure_compare_FE} and \ref{clampedbeam_pressure_Halfspace_KM}, respectively. For $a/h >1$, our assumption that the displacement of the beam's bottom surface may be approximated through Euler-Bernoulli beam theory breaks down. This causes the semi-analytical results to deviate from those of FE computations in Fig.~\ref{clampedbeam_pressure_compare_FE}.

We have followed  \cite{Keer1983smooth} in plotting $a p \left( \bar{x} \right) / P$ along the vertical axis in Figs.~\ref{clampedbeam_pressure_compare_FE}(b) and \ref{clampedbeam_pressure_Halfspace_KM}. A match employing this scale may not guarantee a correspondence of the actual pressure profiles $\varphi \left( \bar{x} \right)$ or $p \left( \bar{x} \right)$. This is because the total load $P$ in the denominator of $a p \left( \bar{x} \right) / P$ is calculated by integrating $\varphi \left( \bar{x} \right)$ in the numerator. Thus, even if a constant factor is missed in $p \left( \bar{x} \right)$, the ratio $a p \left( \bar{x} \right) /P$ will remain unaffected. Given this, the comparison of  pressure profiles shown in Fig.~\ref{clampedbeam_pressure_compare_FE}(a) is more illuminating. Finally, the pressure profiles in Figs.~\ref{clampedbeam_pressure_compare_FE}(b) and \ref{clampedbeam_pressure_Halfspace_KM} do not vary much with the slenderness ration $l/h$. These plots may therefore be utilized to estimate pressures at other $l/h$ as well. 

\begin{figure}[htbp]
\begin{center}
\subfloat[][]{\includegraphics[width=0.5\linewidth]{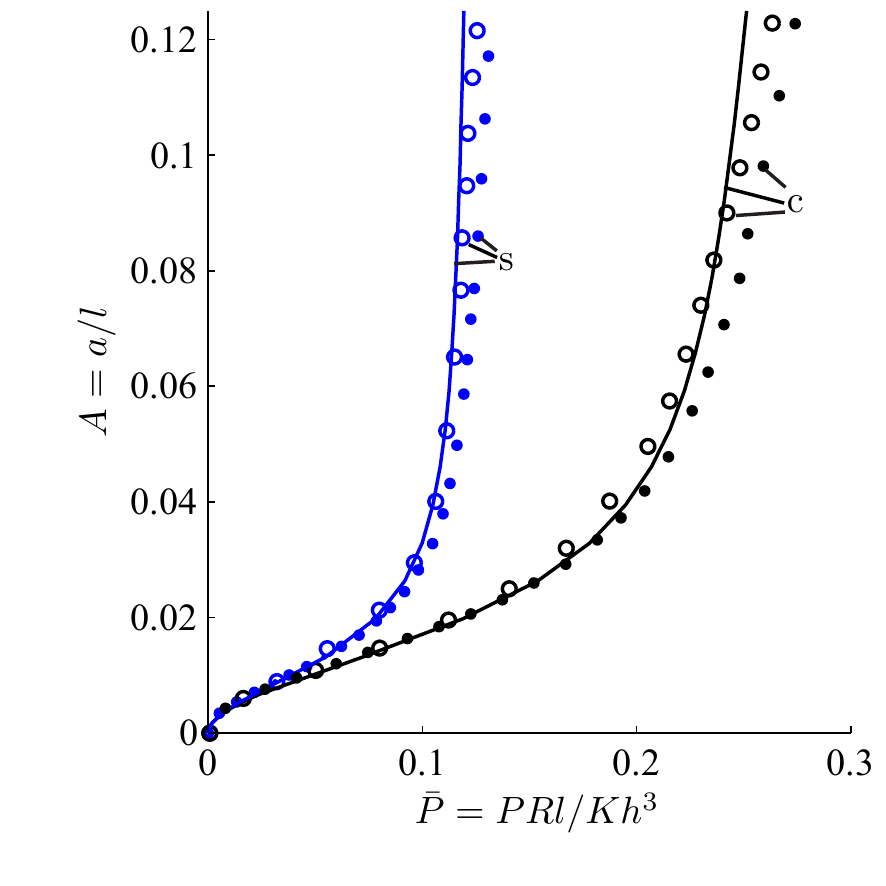}}
\subfloat[][]{\includegraphics[width=0.5\linewidth]{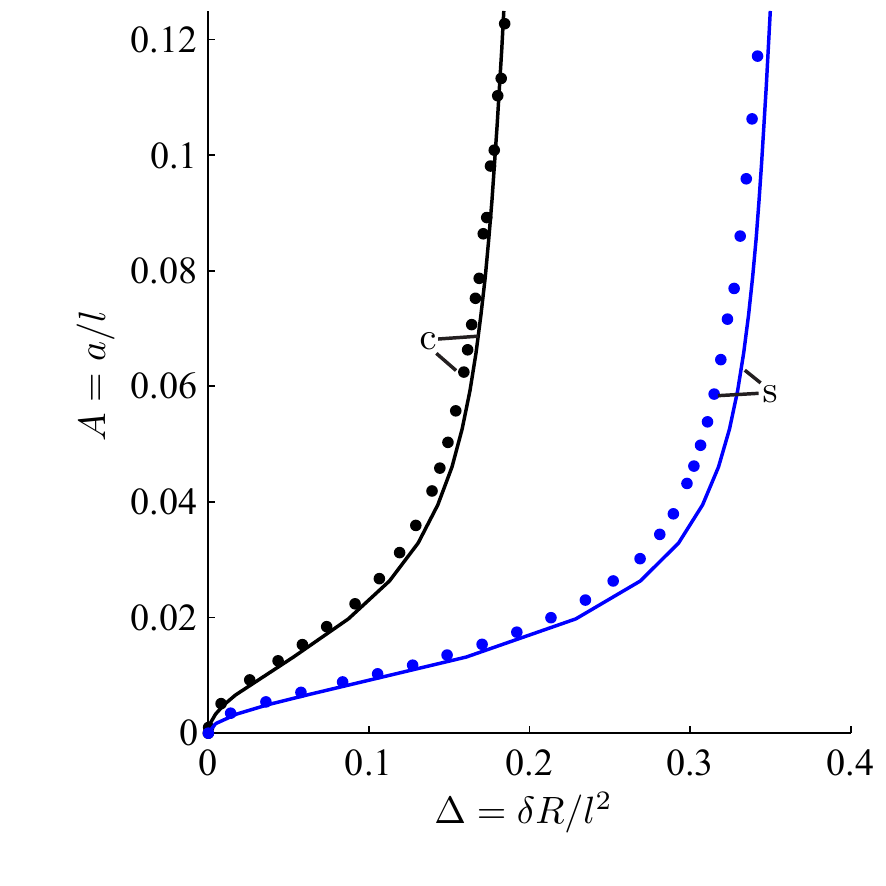}}
\caption{ Non-adhesive contact of clamped (`c') and simply supported (`s') beams. The contact area $A$ is plotted as a function of (a) the total load $\bar{P}$ acting on the punch and (b) the punch's displacement $\Delta$. The beam's slenderness ratio $l/h$=10. Solid lines are results obtained from the semi-analytical procedure of Sec.~\ref{sec:Numerical_solution}. Filled circles correspond to FE simulations of Sec.~\ref{sec:FE_model}. Predictions of \cite{Sankar1983} are shown by open circles, when available.}
\label{fig:fixedbeam_A_Hertz_ABAQUS}
\end{center}
\end{figure}

\begin{figure}[htbp]
\begin{center}
\includegraphics[width=0.5\linewidth]{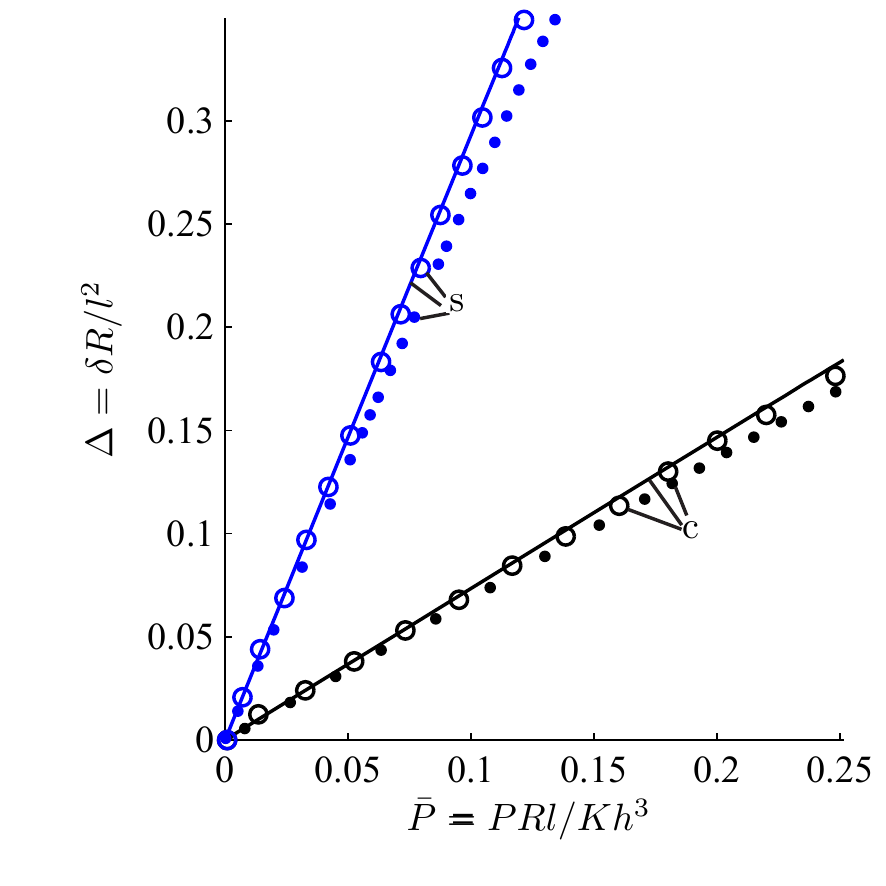}
\caption{Non-adhesive contact of clamped (`c') and simply supported (`s') beams. The displacement $\Delta$ of the punch is shown as a function of the total load $\bar{P}$. See also the caption of Fig.~\ref{fig:fixedbeam_A_Hertz_ABAQUS}.}
\label{fig:fixedbeam_Delta_P_Hertz_ABAQUS}
\end{center}
\end{figure}

Next, the variation of the contact area $A$ with the total load $\bar{P}$ and with the displacement $\Delta$ of the punch are shown in Fig.~\ref{fig:fixedbeam_A_Hertz_ABAQUS}. Results of both clamped and simply supported beams (discussed in the next section) are shown. From Fig.~\ref{fig:fixedbeam_A_Hertz_ABAQUS}(a) we find that  to obtain the same contact area $A$, a clamped beam requires higher load compared to the simply supported beam. Similarly, Fig.~\ref{fig:fixedbeam_A_Hertz_ABAQUS}(b) shows that, at a given $A$, a clamped beam displaces less than a simply supported beam. Both these outcomes are expected, as the bending stiffness of the clamped beam is higher than that of a simply supported beam. Thus, the contribution to vertical displacement $\Delta$ from the beam's bending is lowered in the case of a clamped beam. Similarly, a clamped beam wraps less about the punch, thereby lowering the contact area at given load. The clamped beam's greater bending stiffness compared to that of a simply supported beam is clearly demonstrated by Fig.~\ref{fig:fixedbeam_Delta_P_Hertz_ABAQUS} that plots the deflection of the beam's center point -- which equals the punch's displacement $\Delta$ -- against the total load $\bar{P}$. The linear response of $\Delta$ with $\bar{P}$ is not unexpected as the displacement of the beam's bottom surface  is obtained from beam theory. From Figs.~\ref{fig:fixedbeam_A_Hertz_ABAQUS} and \ref{fig:fixedbeam_Delta_P_Hertz_ABAQUS} it is evident that end supports have significant bearing on the beam's indentation. Finally, in Figs.~\ref{fig:fixedbeam_A_Hertz_ABAQUS} and \ref{fig:fixedbeam_Delta_P_Hertz_ABAQUS}, we again find a good match both  with FE simulations and with the results of \cite{Sankar1983}.

\begin{figure}[htbp]
\centering
\subfloat[][]{\includegraphics[width=0.5\linewidth]{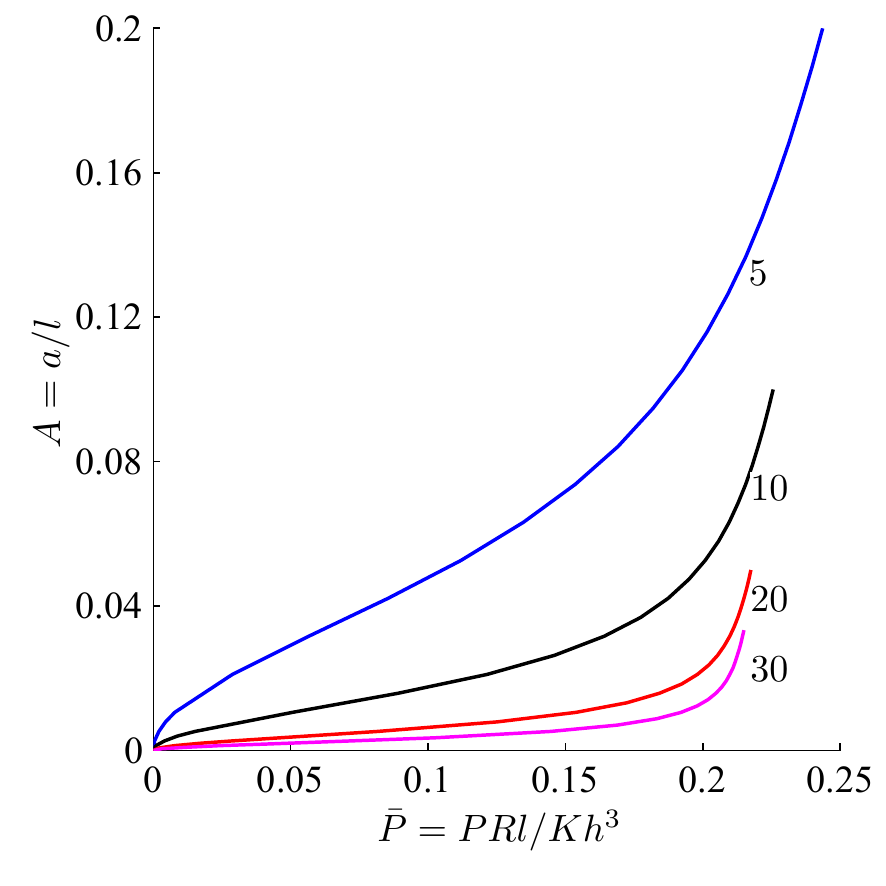}} 
\subfloat[][]{\includegraphics[width=0.5\linewidth]{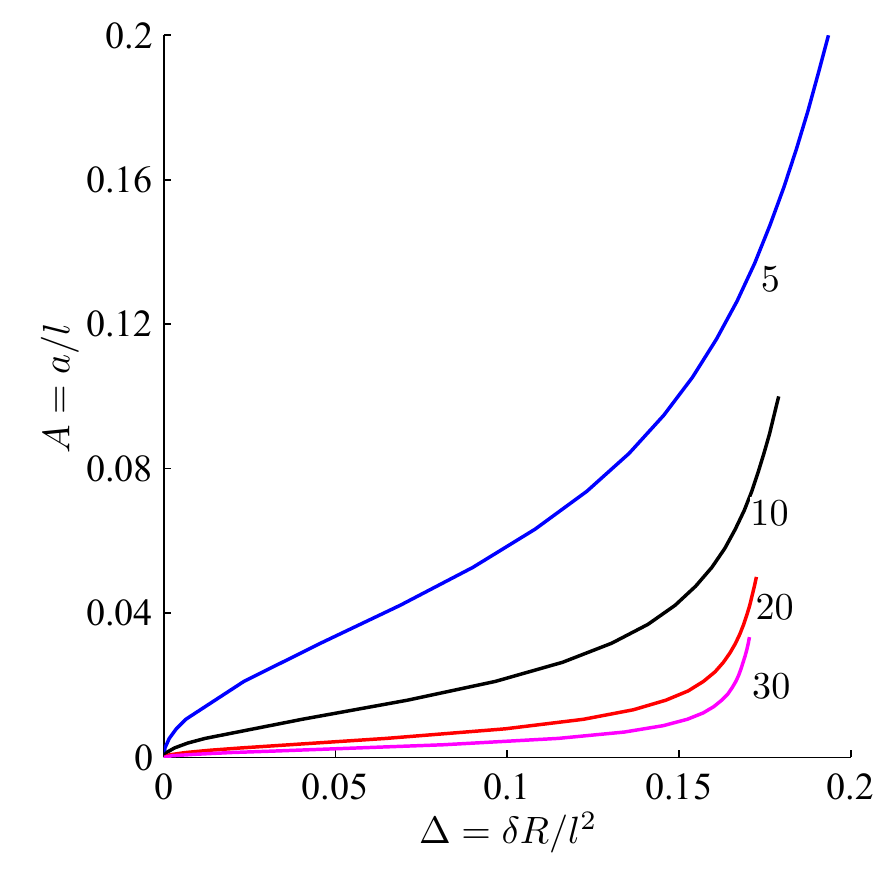}}
\caption{Non-adhesive contact of clamped beams. Variation of contact area $A$ with (a) total load $\bar{P}$ and (b) punch's displacement $\Delta$ is shown. Different slenderness ratios $l/h$ are considered and these are noted next to their associated curves. }
\label{fixed_beam_hertz_diff_lbyh_plots}
\end{figure}

\begin{figure}[htbp]
\centering
{\includegraphics[width=0.5\linewidth]{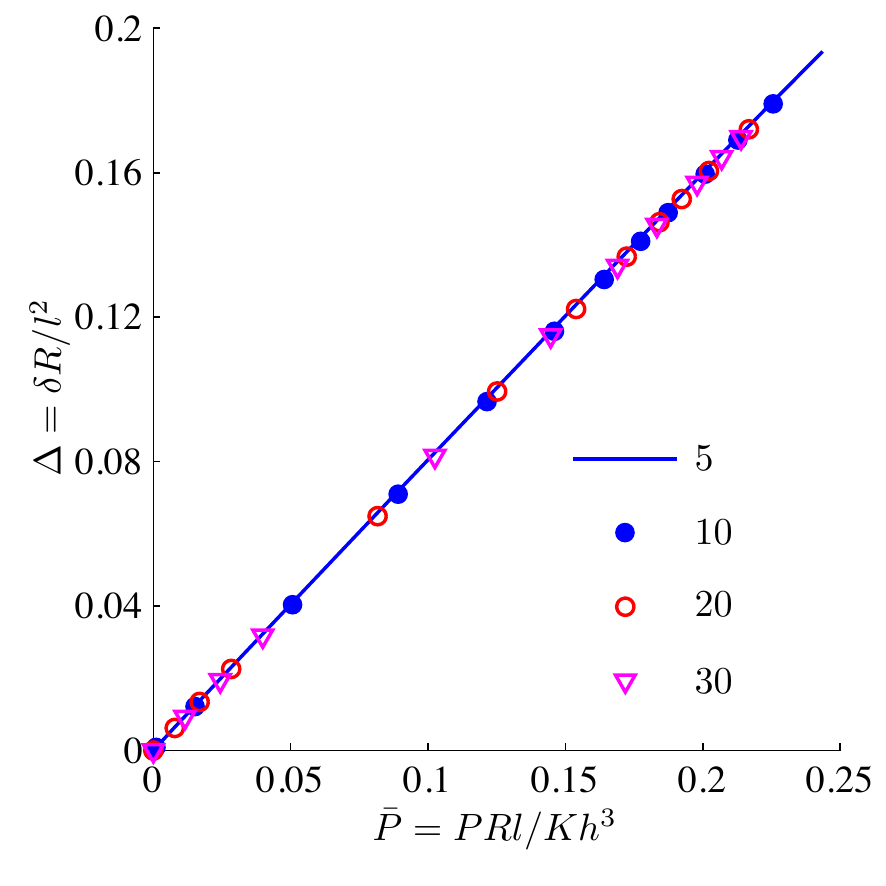}}
\caption{Non-adhesive contact of clamped beams. Variation of punch's displacement $\Delta$ with total load $\bar{P}$ is shown. Different slenderness ratios $l/h$ are considered as noted in the legend. }
\label{fixed_beam_hertz_P_delta_diff_lbyh_plots}
\end{figure}

As mentioned in Sec.~\ref{sec:FE_model}, the material parameters employed in FE simulations may not be relevant for typical applications. Thus, in Figs.~\ref{fixed_beam_hertz_diff_lbyh_plots} and \ref{fixed_beam_hertz_P_delta_diff_lbyh_plots} we report results with parameters more commonly encountered. Following \cite{Dalmeya2012contact}, the Young's modulus and Poisson's ratio of the beam are taken corresponding to those observed in soft materials: $E=0.083$ MPa and $\nu=0.4$, respectively. The beam's geometry remains the same as before. 

The contact area $A$ is plotted against the total load acting on the punch $\bar{P}$ and the displacement $\Delta$ of the punch in Fig.~\ref{fixed_beam_hertz_diff_lbyh_plots} for clamped beams of different slenderness ratios $l/h$. A more slender beam is less resistant to bending. Thus, beams with higher $l/h$ require less load compared to lower $l/h$ beams to achieve the same contact area.  Hence, the curves in Fig.~\ref{fixed_beam_hertz_diff_lbyh_plots}(a) move downward with  increasing slenderness ratio. Similarly, the punch's displacement $\Delta$ is high for more slender beams. Therefore, the curves in Fig.~\ref{fixed_beam_hertz_diff_lbyh_plots}(b) also shift downward with growing $l/h$.

Finally,  in Fig.~\ref{fixed_beam_hertz_P_delta_diff_lbyh_plots} we plot variation of the displacement $\Delta$ of the punch  with the total load $\bar{P}$ for different $l/h$ ratios. We observe that, our choice of non-dimensionalization (cf. Sec.~\ref{sec:Non-dimensionalization}) for the punch's displacement and the total load allows the curves in Fig.~\ref{fixed_beam_hertz_P_delta_diff_lbyh_plots} to collapse onto a single line. This is not seen for other scalings, cf. Secs.~\ref{sec:Results_JKR} and \ref{sec:Results_Maugis}, where we report results for adhesive contact.

\subsection{Effect of end conditions}
\label{Subsec: Effect of end conditions}
We obtain results for a simply supported beam in the limit $k_s^f  \rightarrow \infty$ and $k_t^f \rightarrow 0$. The vertical displacement of the bottom surface is given by \eqref{V_omega_defn}. The contact pressure $\varphi \left( \bar{x} \right)$ is then found by solving \eqref{Int_eqn_Hertz} and \eqref{end_pressure}, and invoking \eqref{phi_cheby}. The behavior of a simply supported beam is qualitatively similar to that of a clamped beam, but differs quantitatively.

\begin{figure}[htbp]
\begin{center}
\subfloat[][]{\includegraphics[width=0.5\linewidth]{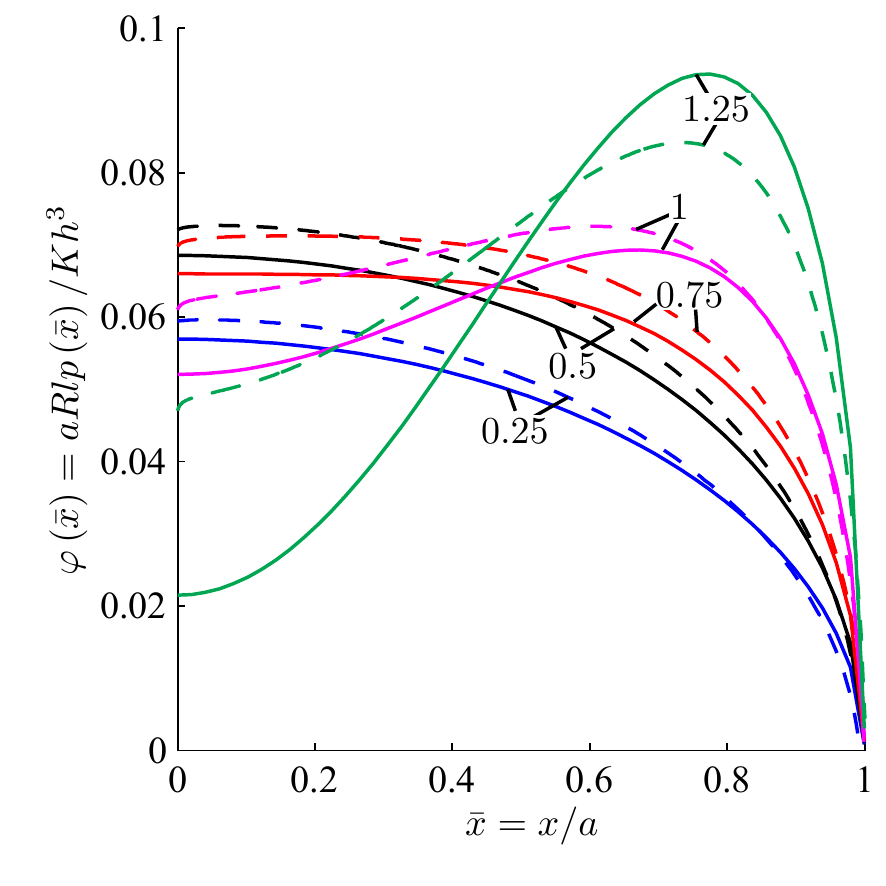}}
\subfloat[][]{\includegraphics[width=0.5\linewidth]{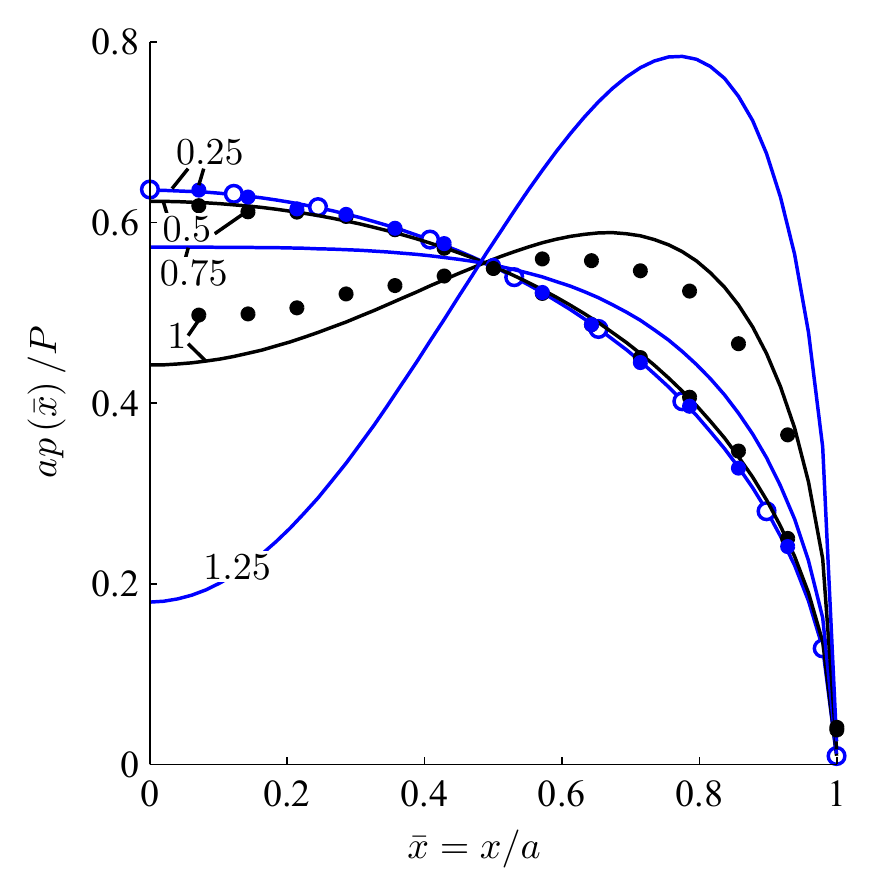}}
\caption{Non-adhesive contact of a simply supported beam. Non-dimensional contact pressure (a) $\varphi \left( \bar{x} \right)$ and (b) $a p \left( \bar{x} \right) / P$ are shown. Several contact areas are investigated by varying $a/h$ as noted next to the associated curves, while keeping $l/h=10$. Solid lines are results obtained from the semi-analytical procedure of Sec.~\ref{sec:Numerical_solution}. Dashed lines in (a) correspond to FE simulations of Sec.~\ref{sec:FE_model}. Open circles in (b) represent the solution for an elastic half-space. Results of \cite{Keer1983smooth}, when available, are shown in (b) by filled circles.} 
\label{simplebeam_pressure_compare_FE} 
\end{center}
\end{figure}

Figure~\ref{simplebeam_pressure_compare_FE} repeats Figs.~\ref{clampedbeam_pressure_compare_FE} and \ref{clampedbeam_pressure_Halfspace_KM} for simply supported beams and compare our semi-analytical results with those of FE simulations, and \cite{Keer1983smooth}. As before, we find good agreement between all three approaches for $a/h \lessapprox 1$.

Contrasting Figs.~\ref{clampedbeam_pressure_compare_FE}(a) and \ref{simplebeam_pressure_compare_FE}(a) we find that, at the same $a/h$, pressures found in a simply supported beam are lower compared to those in a clamped beam. Thus, the total load $\bar{P}$ required to achieve the same contact area, for clamped and simply supported beams of the same thickness, is very different. This reinforces the importance of  correctly modeling end supports in beam indentation. Interestingly, because of the manner in which the contact pressure is scaled, Figs.~\ref{clampedbeam_pressure_Halfspace_KM} and \ref{simplebeam_pressure_compare_FE}(b) are nearly the same.

\begin{figure}[htbp]
\begin{center}
\subfloat[][]{\includegraphics[width=0.46\linewidth]{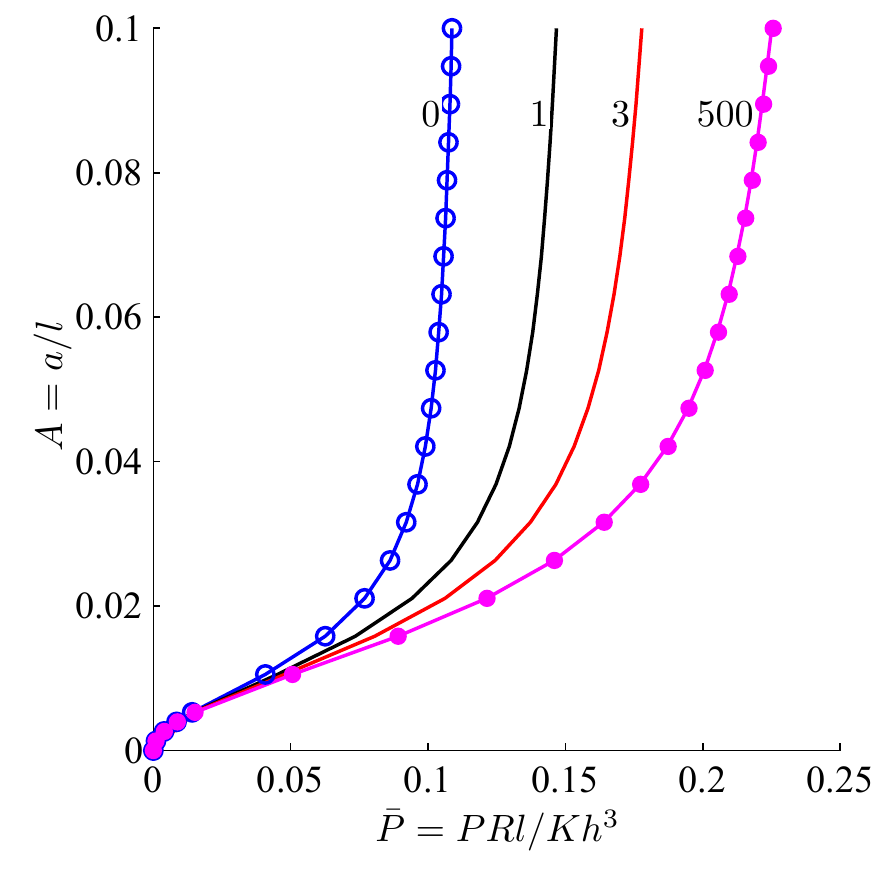}} 
\subfloat[][]{\includegraphics[width=0.46\linewidth]{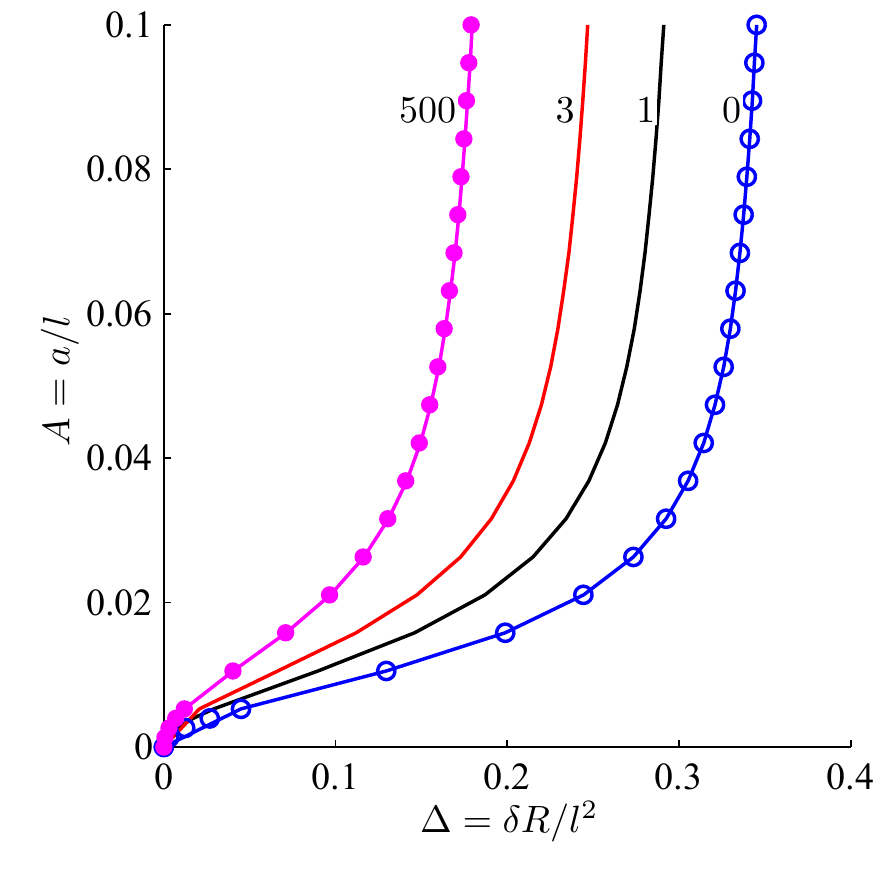}}  
\caption{Non adhesive contact of beams on flexible supports. The contact area $A$ is plotted as a function of (a) the total load $\bar{P}$ acting on the punch and (b) the punch's displacement $\Delta$. The beam's slenderness ratio $l/h=10$. The vertical translational spring's stiffness $k_s^f=\infty$. Various torsional springs are considered and their stiffnesses $k_t^f$ are indicated next to their associated curves. Open and filled circles represent results for simply supported and a clamped beams, respectively.}
\label{flexiblebeam_hertz_diff_ktf_plots}
\end{center}
\end{figure}

\begin{figure}[htbp]
\centering
\subfloat[][]{\includegraphics[width=0.5\linewidth]{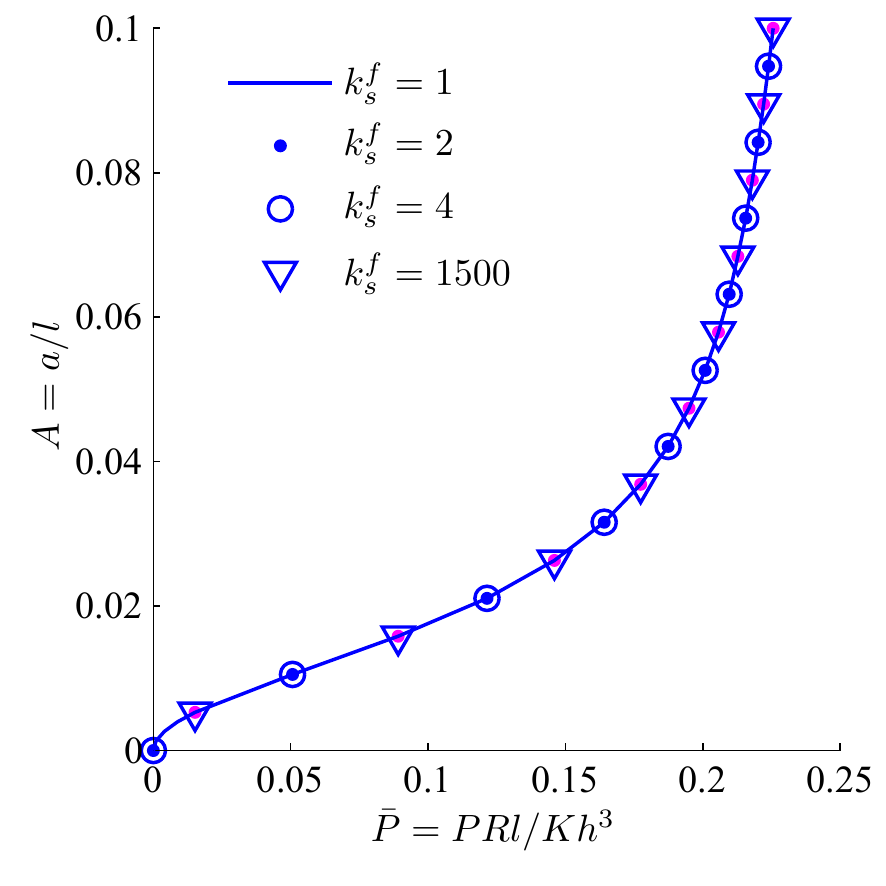}} 
\subfloat[][]{\includegraphics[width=0.5\linewidth]{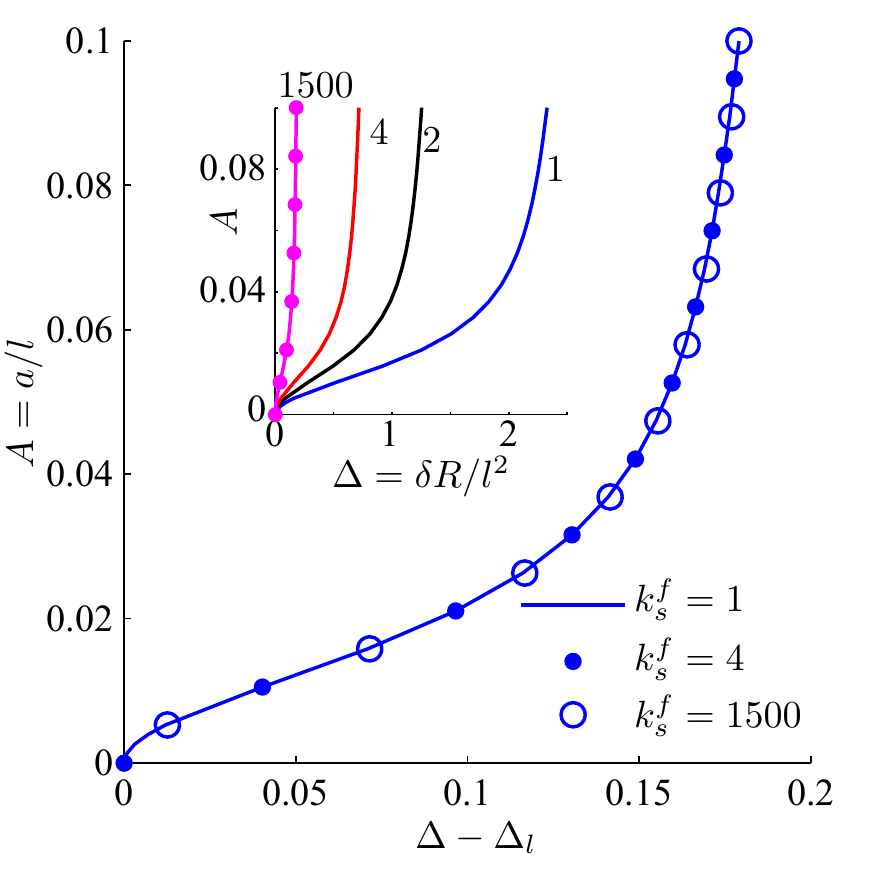}} 
\caption{Non-adhesive contact of beams on flexible supports. The contact area $A$ is plotted as a function of (a) the total load $\bar{P}$ acting on the punch and (b) the adjusted punch displacement $\Delta-\Delta_l$.  The inset in (b) shows the variation of the contact area $A$ with the punch's displacement $\Delta$. The beam's slenderness ratio $l/h=10$. The torsional spring stiffness $k_t^f=\infty$. Various vertical translational springs are considered and their stiffnesses $k_s^f$ are indicated either in the legend or next to their associated curves. Filled circles in the inset in (b) represent results for a clamped beam.}
\label{flexiblebeam_hertz_diff_ksf_plots}
\end{figure}

Finally, we report results on the non-adhesive contact of beams resting on flexible supports with parameters utilized to generate Figs.~\ref{fixed_beam_hertz_diff_lbyh_plots} and \ref{fixed_beam_hertz_P_delta_diff_lbyh_plots}. Again, for beams with slenderness ratio $l/h=10$, the contact area $A$ is plotted against the total load $\bar{P}$ acting on the punch and the punch's displacement $\Delta$ in Figs.~\ref{flexiblebeam_hertz_diff_ktf_plots} and \ref{flexiblebeam_hertz_diff_ksf_plots}. Figure~\ref{flexiblebeam_hertz_diff_ktf_plots} shows results for several torsional spring stiffnesses $k_t^f$ after setting the vertical translational spring's stiffness $k_s^f$ to infinity. Such a beam may be thought of as a simply supported beam with some resistance to rotation at the ends, or a beam whose clamped ends allow some rotational play. Results lie between those obtained for clamped and simply supported beams. Expectedly, increasing $k_t^f$ shifts the results towards those of a clamped beam, and decreasing it yields results close to those of a simply supported beam. This is seen clearly in Fig.~\ref{flexiblebeam_hertz_diff_ktf_plots}.

Figure~\ref{flexiblebeam_hertz_diff_ksf_plots} repeats Fig.~\ref{flexiblebeam_hertz_diff_ktf_plots}, but this time keeping $k_t^f$ as infinity and varying $k_s^f$. We find that increasing $k_s^f$ does \emph{not} affect the variation of $A$ with $\bar{P}$, but the dependence of $A$  on $\Delta$ changes; see the inset in Fig.
~\ref{flexiblebeam_hertz_diff_ksf_plots}(b). The latter change is, however, due to the vertical displacement $\Delta_l$ of the beam's spring supports. The presence of $\Delta_l$ shifts the datum downwards, so that indentation now initiates from $y=\Delta_l$, rather than from $y=0$. Once we correct for $\Delta_l$ we find that displacement plots in the inset of Fig.~\ref{flexiblebeam_hertz_diff_ksf_plots}(b) are also unaffected by variation in $k_s^f$, as shown in Fig.~\ref{flexiblebeam_hertz_diff_ksf_plots}(b).

\section{Results: Adhesive contact - JKR approximation}
\label{sec:Results_JKR}
We now consider adhesive contact of beams after invoking the JKR approximation, previously discussed in Secs.~\ref{sec:Mathematical_model} and \ref{sec:Non-dimensionalization}. Thus, we need to solve the integral equation \eqref{Int_eqn_final} along with energy balance \eqref{fracture_Griffith_nondim}, in the limit of adhesive strength dominating elastic stiffness, i.e. $\lambda \rightarrow \infty$, while the adhesive zone becomes infinitesimally small, so that $\bar{c} \rightarrow 1$. Employing expansion \eqref{phi_cheby} for the contact pressure,  \eqref{Int_eqn_final} and \eqref{fracture_Griffith_nondim} become, respectively,
\begin{equation}
\sum_{n=0}^{N} a_{2n} \mathcal{J}_{nm} = \frac{8 \hat{\gamma}^3}{3 \pi} \sum_{n=0}^{N} b_{2n} \mathcal{J}^{\left(1\right)}_{nm} + \frac{4 b_0 }{3 \hat{\gamma} \bar{I} \left( 1-\nu^2 \right)} \mathcal{J}^{\left(2\right)}_{m} 
\quad \text{for } m=0,\cdots, N
\label{Int_eqn_JKR}
\end{equation} 
and
\begin{equation}
\label{end_pressure_JKR}
b_{0} + b_{2} + \cdots + b_{2N} = -\frac{m}{2 \pi L} \left( \frac{l}{h} \right)^3 \sqrt{\frac{6 A m}{L}} .
\end{equation}

From here on, we follow \cite{Maugis1992adhesion} and employ \begin{equation}
\hat{A} = \frac{AL}{m} = a \left( \frac{K}{\pi w R^2} \right)^{1/3}, \quad \hat{P} = \frac{\bar{P} H^3}{L m^3} = \frac{P}{\pi w} \quad \text{ and} \quad \hat{\Delta} = \frac{\Delta L^2}{m^2} =  \delta \left( \frac{K^2}{\pi^2 w^2 R} \right)^{1/3}, 
\end{equation}
where $H=h/R$, instead of, respectively, $A$, $P$, and $\Delta$, to report our results. This is done in order to facilitate contact with other work on adhesion. We set the adhesion energy $w=0.02 \times 10^{-3} \text{ J}/\text{mm}^2$.

\begin{figure}[htbp]
\centering
\subfloat[][]{\includegraphics[width=0.5\linewidth]{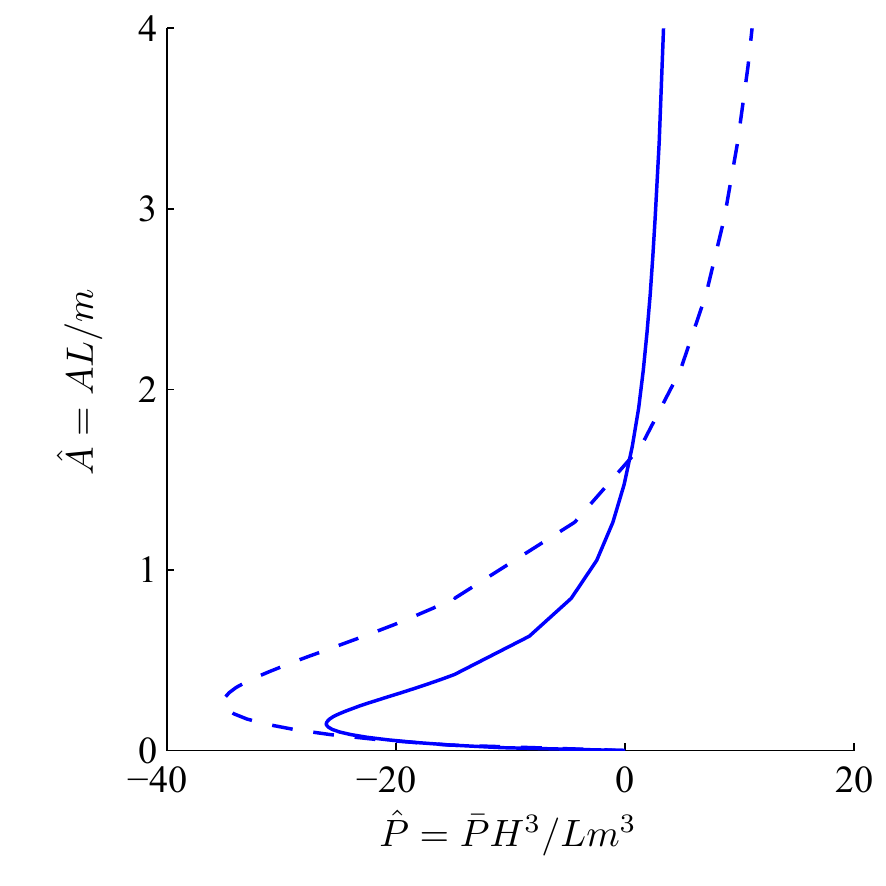}} 
\subfloat[][]{\includegraphics[width=0.5\linewidth]{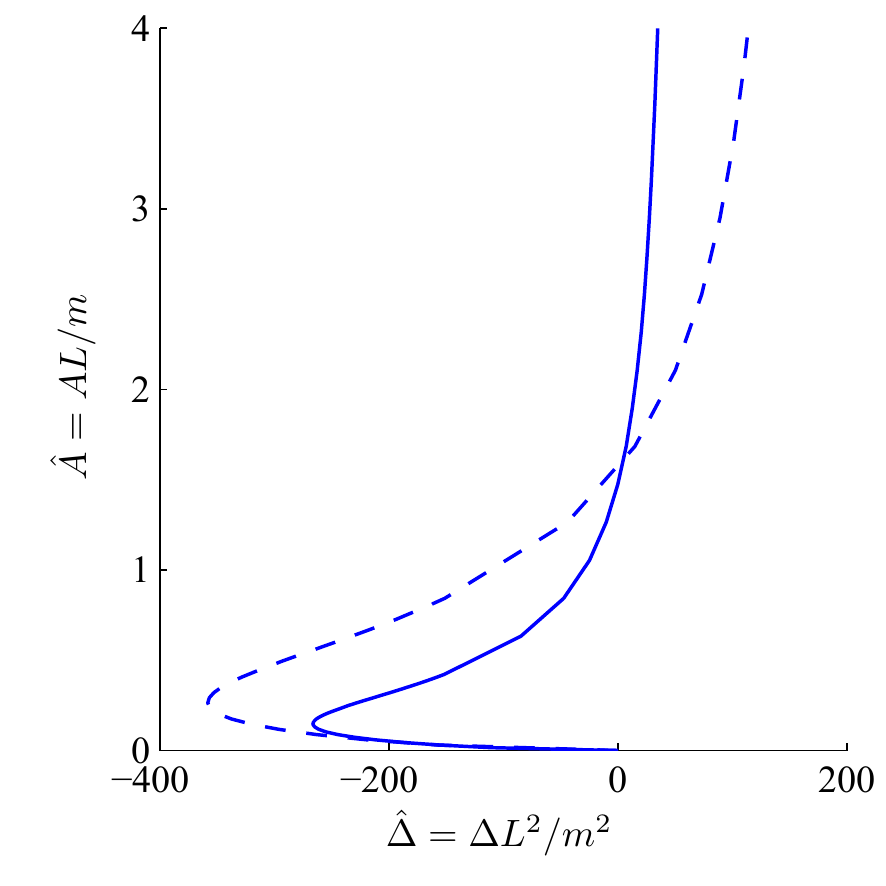}}
\caption{Adhesive contact of clamped beams with the JKR approximation. Variation of contact area $\hat{A}$ with the (a) total load $\hat{P}$ and (b) the punch's displacement $\hat{\Delta}$. The beam's slenderness ratio $l/h=10$.  Solid lines correspond to $l=40$ mm and $h=4$ mm, while the dashed line is for a beam with $l=80$ mm and $h=8$ mm.}
\label{fixedbeam_jkr_same_lbyh_plots}
\end{figure}

\begin{figure}[htbp]
\centering
\begin{multicols}{2}
\includegraphics[width=\linewidth]{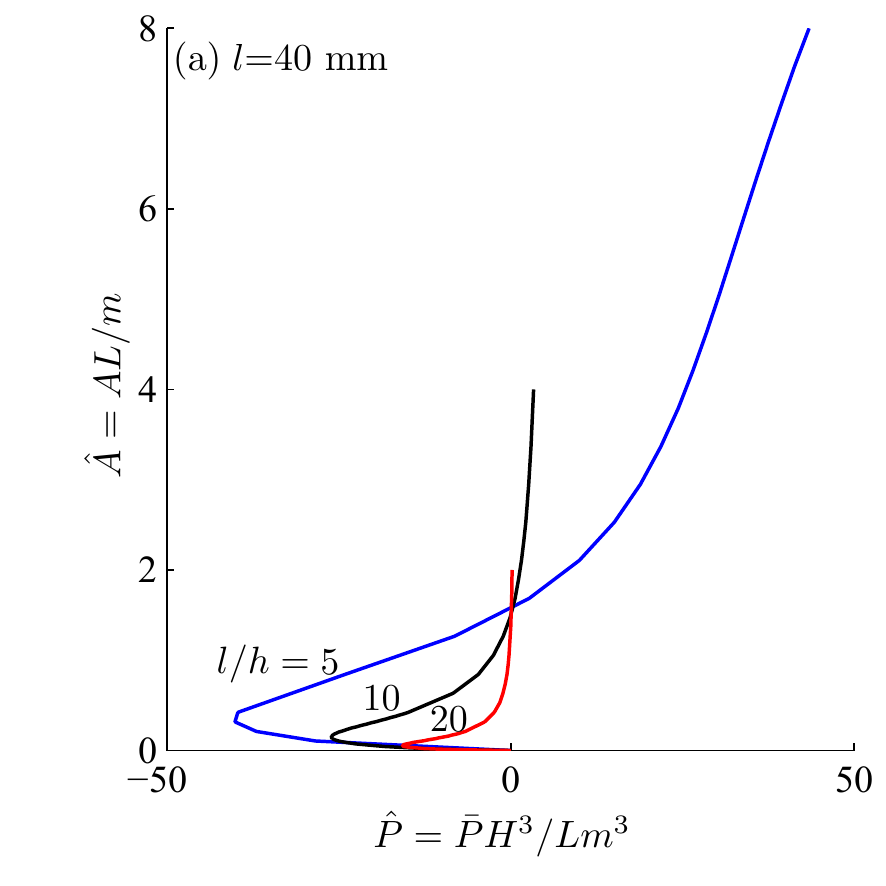} 
\includegraphics[width=\linewidth]{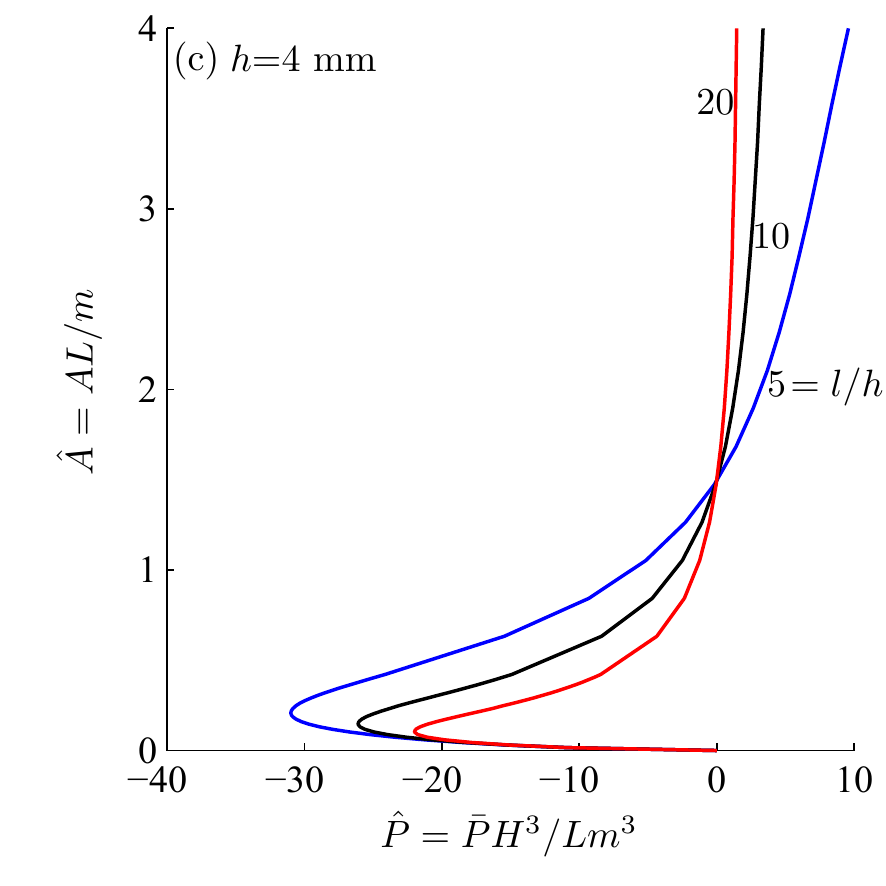}
\includegraphics[width=\linewidth]{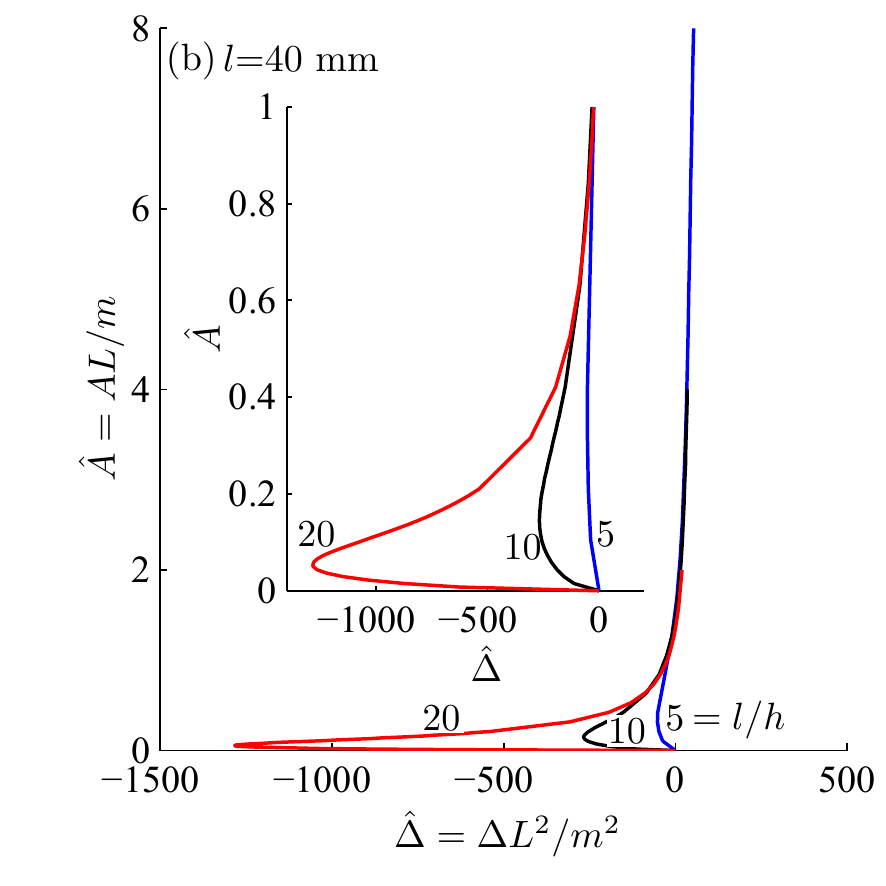} 
\includegraphics[width=\linewidth]{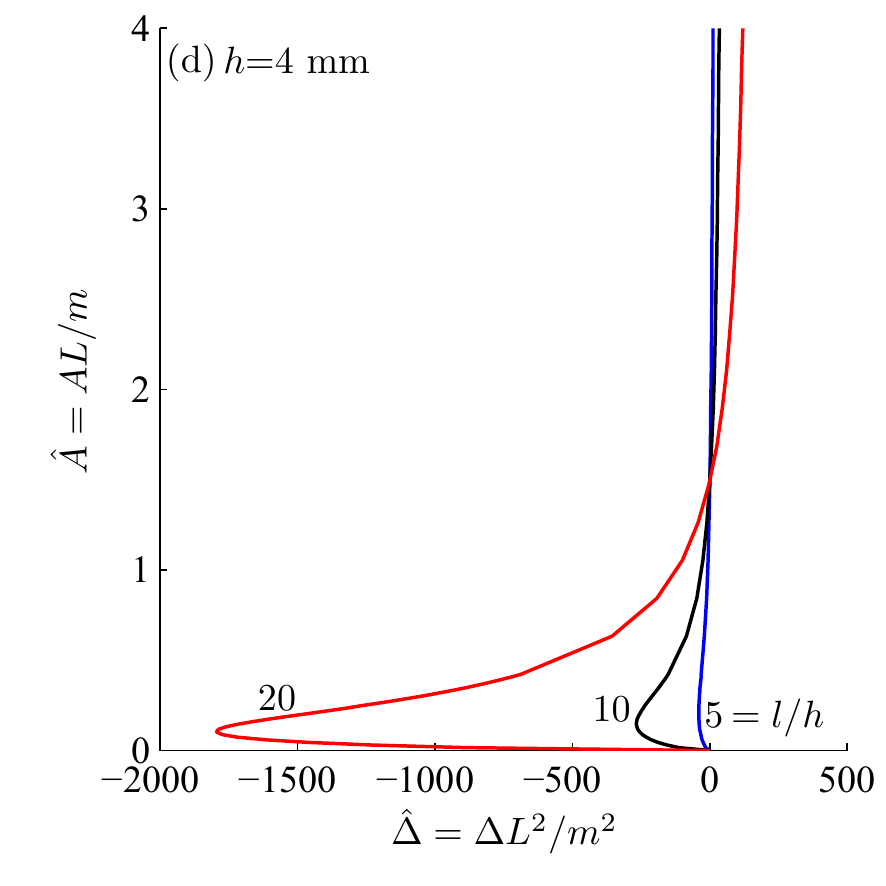}  
\end{multicols}
\caption{Adhesive contact of clamped beams with the JKR approximation.  Left column reports the variation of contact area $\hat{A}$ with total load $\hat{P}$, while the right column plots the change of $\hat{A}$ with the punch's displacement $\hat{\Delta}$. Results in the top row are obtained by setting $l=40$ mm and varying $h$ as shown, while those in the bottom row have  $h=4$ mm but different $l$, as indicated. The inset in (b) illustrates behavior at low $\hat{A}$.}
\label{fixedbeam_jkr_diff_lbyh_plots}
\end{figure}

Figure~\ref{fixedbeam_jkr_same_lbyh_plots} plots the variation of the contact area $\hat{A}$ with the total load $\hat{P}$ acting on the punch and with punch's displacement $\hat{\Delta}$ for clamped beams. While the slenderness ratio $l/h=10$, two different  combinations of $l$ and $h$ are considered. From Fig.~\ref{fixedbeam_jkr_same_lbyh_plots} we observe that the variation of $\hat{A}$  with $\hat{P}$ and $\hat{\Delta}$ is sensitive to the choice of $l$ and $h$, notwithstanding the fact that $l/h$ is kept constant. This is in contrast to the case of non-adhesive contact of Sec.~\ref{sec: Comparision_with_ABAQUS}, where results depended only on $l/h$. However, this behavior is expected for adhesive beams as the right hand side of \eqref{end_pressure_JKR} depends on $L=l/R$. This aspect is further exemplified in Fig.~\ref{fixedbeam_jkr_diff_lbyh_plots}, which plots $\hat{A}$ against $\hat{P}$ and $\hat{\Delta}$ for clamped beams for different choices of $l$ and $h$. 

\begin{figure}[htbp]
\begin{center}
\subfloat[][]{\includegraphics[width=0.5\linewidth]{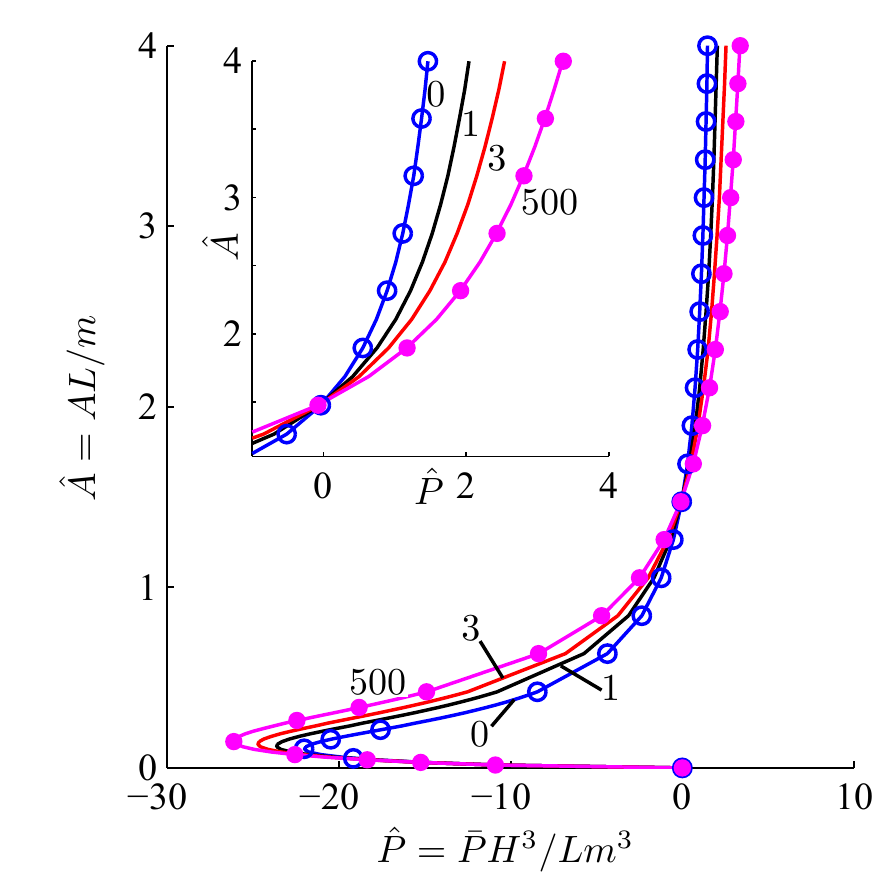}}
\subfloat[][]{\includegraphics[width=0.5\linewidth]{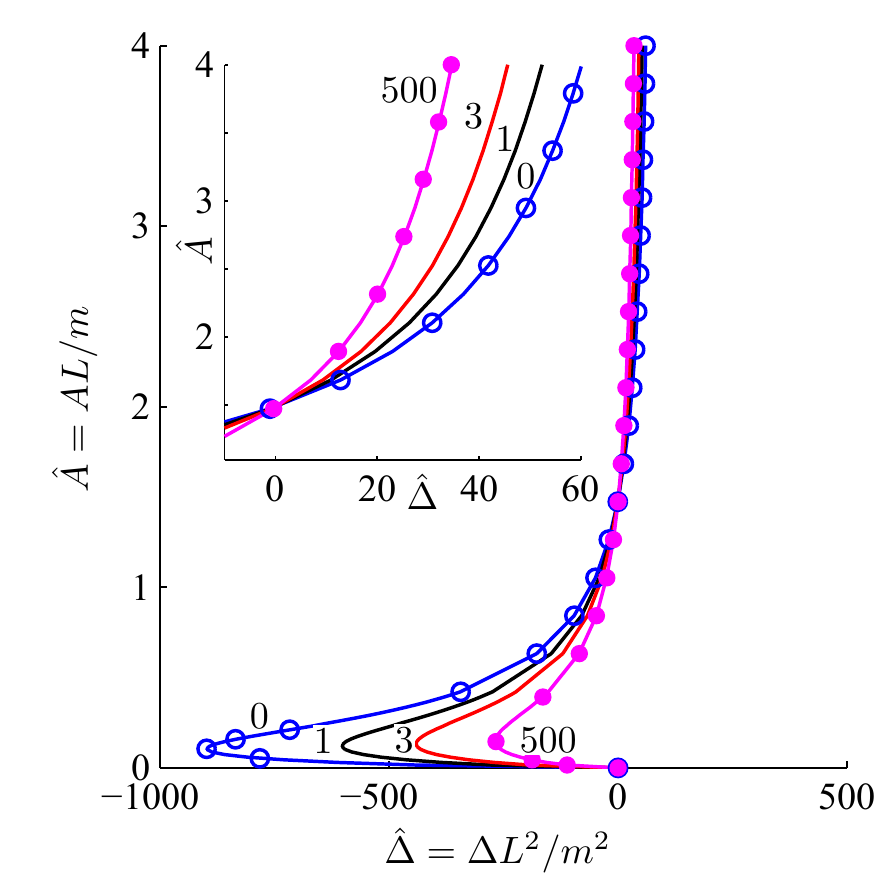}}  
\caption{Adhesive contact of beam on flexible supports with the JKR approximation. The contact area $\hat{A}$ is plotted as a function of (a) the total load $\hat{P}$ acting on the punch and (b) the punch's displacement $\hat{\Delta}$. The beam's thickness $h=4$ mm and $l=40$ mm. The vertical translational spring's stiffness $k_s^f=\infty$. Various torsional springs are considered and their stiffnesses $k_t^f$ are indicated next to their associated curves. The insets correspond to behavior at high $\hat{A}$. Open and filled circles represent results for adhesive beams that are, respectively, simply supported and clamped.}
\label{flexiblebeam_jkr_diff_ktf_plots}
\end{center}
\end{figure}

In Figs.~\ref{fixedbeam_jkr_same_lbyh_plots} and \ref{fixedbeam_jkr_diff_lbyh_plots}, negative values of $\hat{P}$ and $\hat{\Delta}$ reflect tensile loads and upward displacement of the punch, respectively. We recall that $\hat{\Delta}$ equals the deflection of the center point on the beam's top surface, so that $\hat{\Delta}<0$ indicates that the beam bends upwards. Negative values of $\hat{P}$ and $\hat{\Delta}$ are due to the attractive adhesive forces. Due to adhesion, the beam bends upwards and jumps into contact. Equivalently, adhesive forces also act on the punch to pull it down, so that we require a tensile force to hold the punch in it's place.  This tensile force $\hat{P}$ is small for slender beams as they bend easily. For the same reason, this tensile force is smaller for a simply supported beam compared to a clamped beam of the same thickness; cf. Fig.~\ref{flexiblebeam_jkr_diff_ktf_plots}. Once contact is established, the tensile force is slowly released and replaced by a  compressive (downwards) force in order to increase downward indentation. Again, for compressive loads, thin beams bend more easily to wrap around the punch. Thus, slender beams show greater contact area and displacement at the same compressive load $\hat{P}$. This explains the intersection of the curves in Fig.~\ref{fixedbeam_jkr_diff_lbyh_plots}. Therefore, with increasing slenderness ratio, the $\hat{A}$--$\hat{P}$ curves in the left column of Fig.~\ref{fixedbeam_jkr_diff_lbyh_plots} move towards (inwards) the zero-load ($\hat{P}=0$) vertical line, and $\hat{A}$--$\hat{\Delta}$ curves in the right column of Fig.~\ref{fixedbeam_jkr_diff_lbyh_plots} move away (outwards) from the $\hat{\Delta}=0$ line.

Finally, for adhesive beams on flexible supports we plot the contact area $\hat{A}$ against the load $\hat{P}$ and displacement $\hat{\Delta}$ for various  $k_t^f$ in Fig.~\ref{flexiblebeam_jkr_diff_ktf_plots}. Convergence to the results obtained for clamped and simply supported beams may be observed in Fig.~\ref{flexiblebeam_jkr_diff_ktf_plots} by varying $k_t^f$. From our discussion in Sec.~\ref{Subsec: Effect of end conditions}, we know that the variation of $k_s^f$ does not affect how the contact area $\hat{A}$ varies with the load $\hat{P}$. At the same time, change in $\hat{A}$ with the displacement $\hat{\Delta}$ is affected by variation in $k_s^f$ only through the vertical displacement of the translational springs supporting the beam at its ends. By removing this global displacement $\hat{\Delta_l}$ from $\hat{\Delta}$ -- as in Sec.~\ref{Subsec: Effect of end conditions} -- the response of $\hat{A}$ to $\hat{\Delta}$-$\hat{\Delta_l}$ is found to be invariant to $k_s^f$. 

\section{Results: Adhesive contact with an adhesive zone model}
\label{sec:Results_Maugis}
We finally consider contact with an adhesive beam within the framework of adhesive zone models. As already mentioned, we will assume that an adhesive zone of length $d=c-a$ extends outside the contact zone; cf. Fig.~\ref{fig:struct_adhesive_model_text2}(b). Within the adhesive zone the interaction is modeled through the Dugdale-Barenblatt model of \eqref{eqn:Pfn}. To obtain the contact pressure $\varphi \left( \bar{x} \right)$, the displacement $\Delta$ and the location $c$ of the adhesive zone's edge, we have to solve \eqref{Int_eqn_numerical_final}--\eqref{Griffith_Eqn_numerical}. 

\begin{figure}[htbp]
\begin{center}
\subfloat[][]{\includegraphics[width=0.5\linewidth]{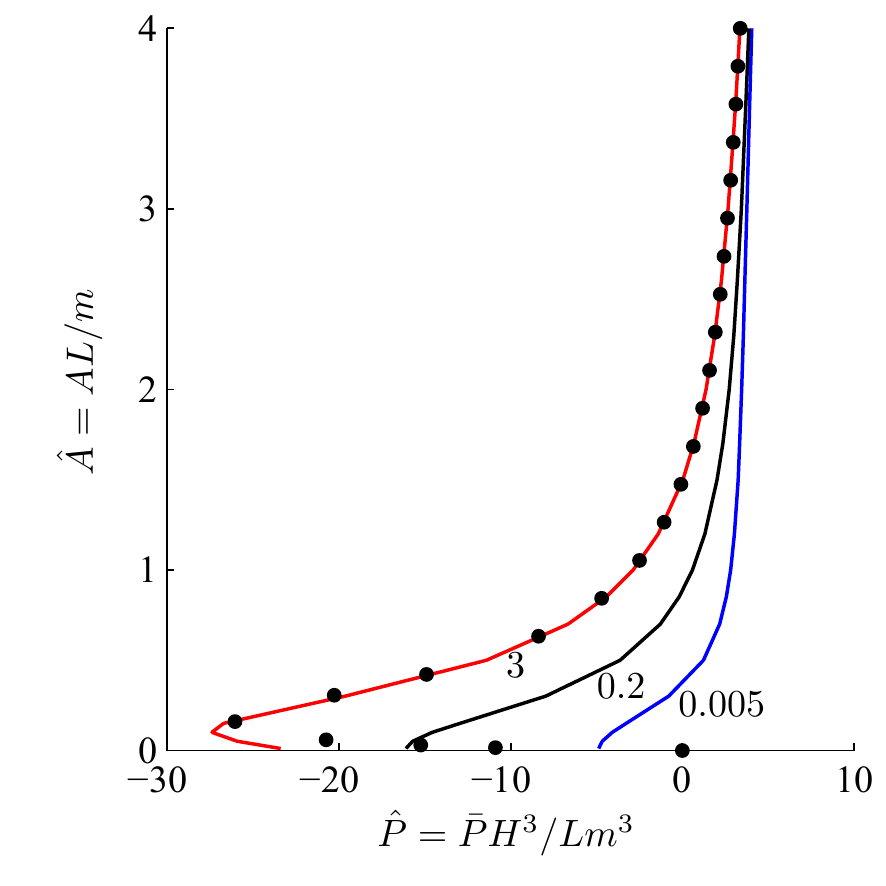}}
\subfloat[][]{\includegraphics[width=0.5\linewidth]{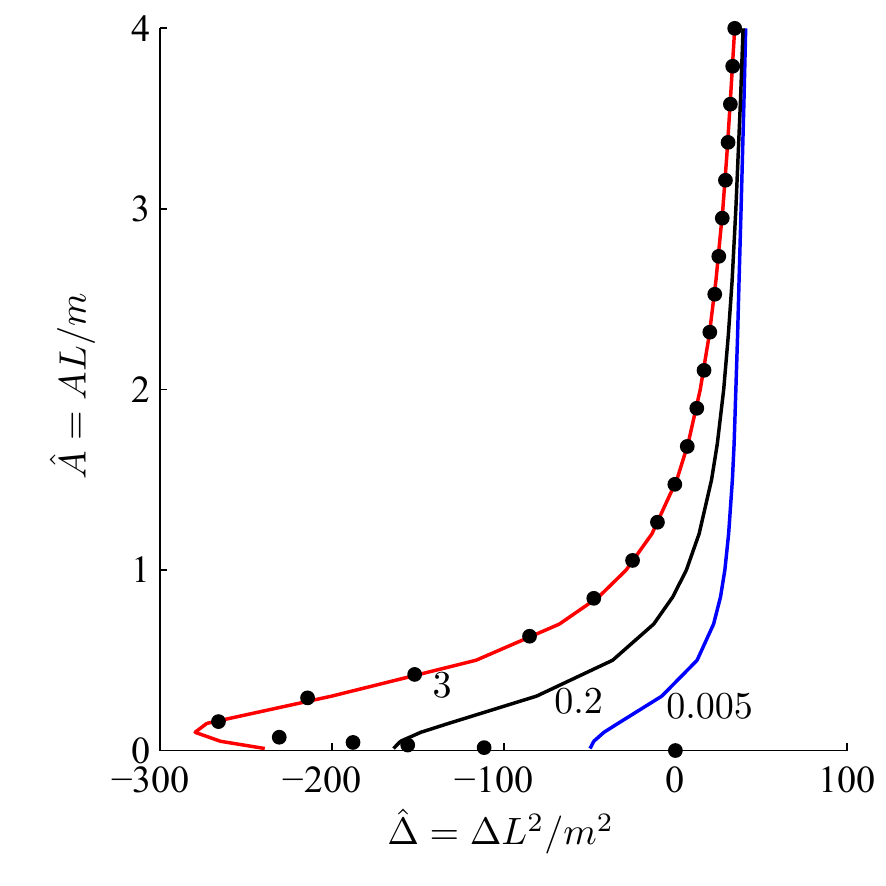}} 
\caption{Adhesive contact of clamped beams with an adhesive zone model. Variation of contact area $\hat{A}$ with (a) the total load $\hat{P}$ and (b) the punch's displacement $\hat{\Delta}$. Different adhesive strengths $\lambda$ are considered and these are indicated next to their associated curves. The beam's thickness $h=4$ mm and $l=40$ mm. Filled circles represent the JKR solution for the corresponding beam; cf. Sec.~\ref{sec:Results_JKR}}
\label{fixedbeam_Maugis_diffF_plots}
\end{center}
\end{figure}

For the clamped beam we plot in Fig.~\ref{fixedbeam_Maugis_diffF_plots} the contact area $\hat{A}$ against the total load $\hat{P}$ and displacement $\hat{\Delta}$ for various adhesive strengths $\lambda$. With increase in $\lambda$, the solutions approach the JKR solution, and we see a close match at $\lambda=3$. On the other hand, as $\lambda \rightarrow 0$, i.e. as adhesion reduces, solutions approach those obtained for non-adhesive contact in Sec.~\ref{sec: Comparision_with_ABAQUS}. 

\begin{figure}[htbp]
\begin{center}
\begin{multicols}{2}
\includegraphics[width=\linewidth]{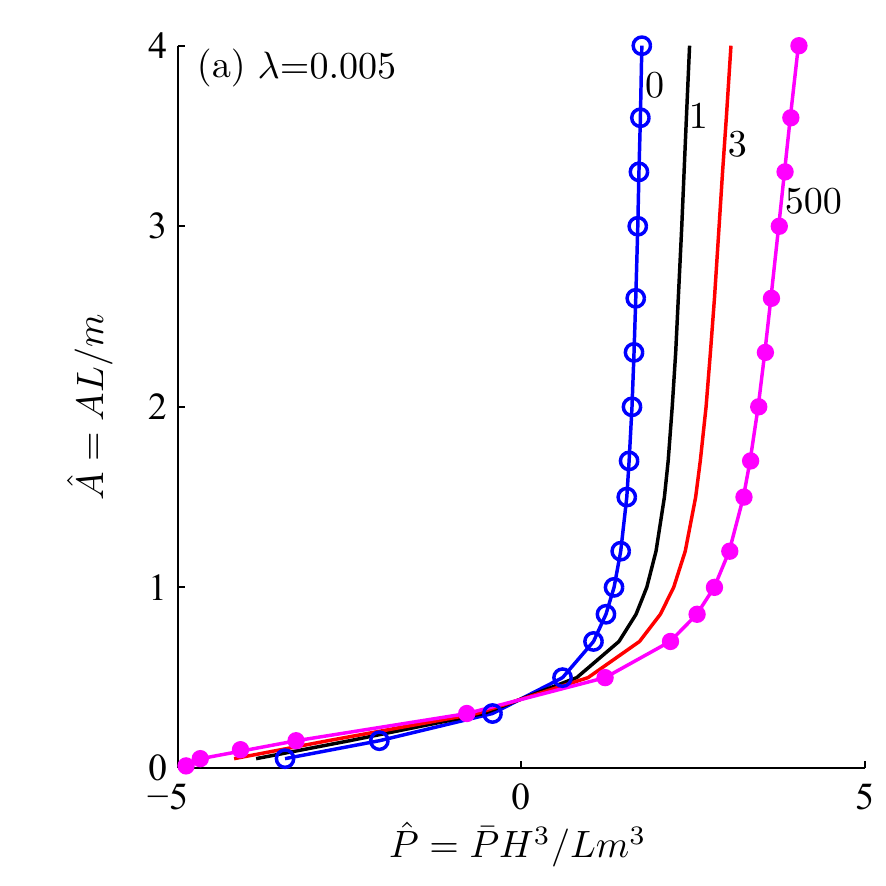} 
\includegraphics[width=\linewidth]{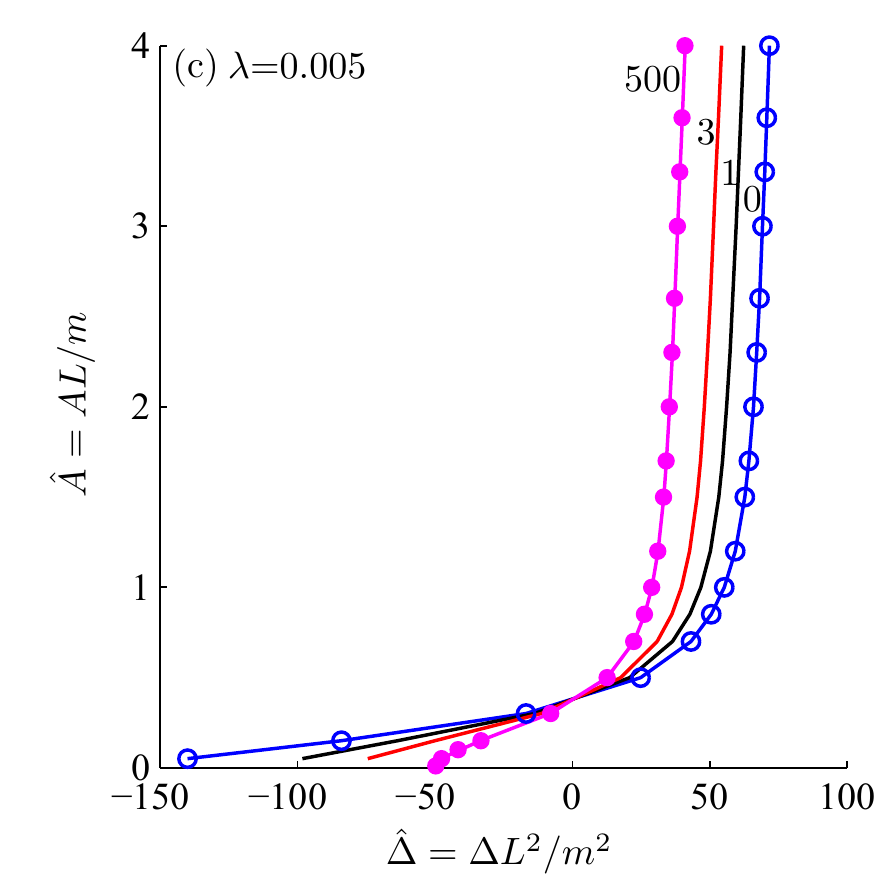}  
\includegraphics[width=\linewidth]{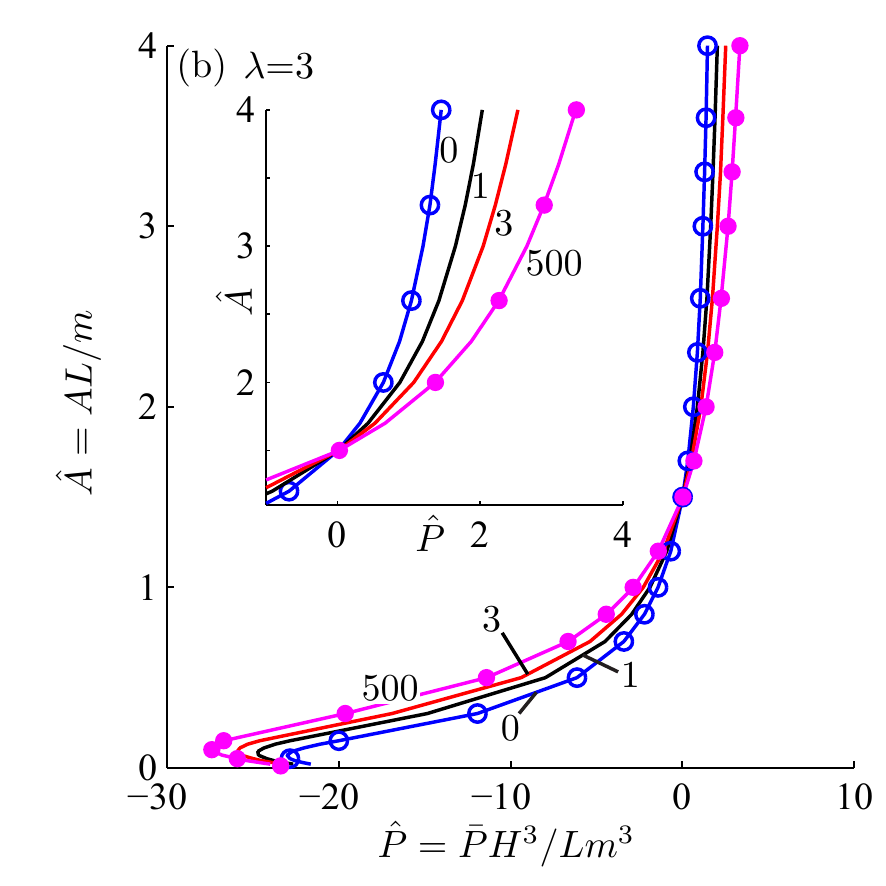} 
\includegraphics[width=\linewidth]{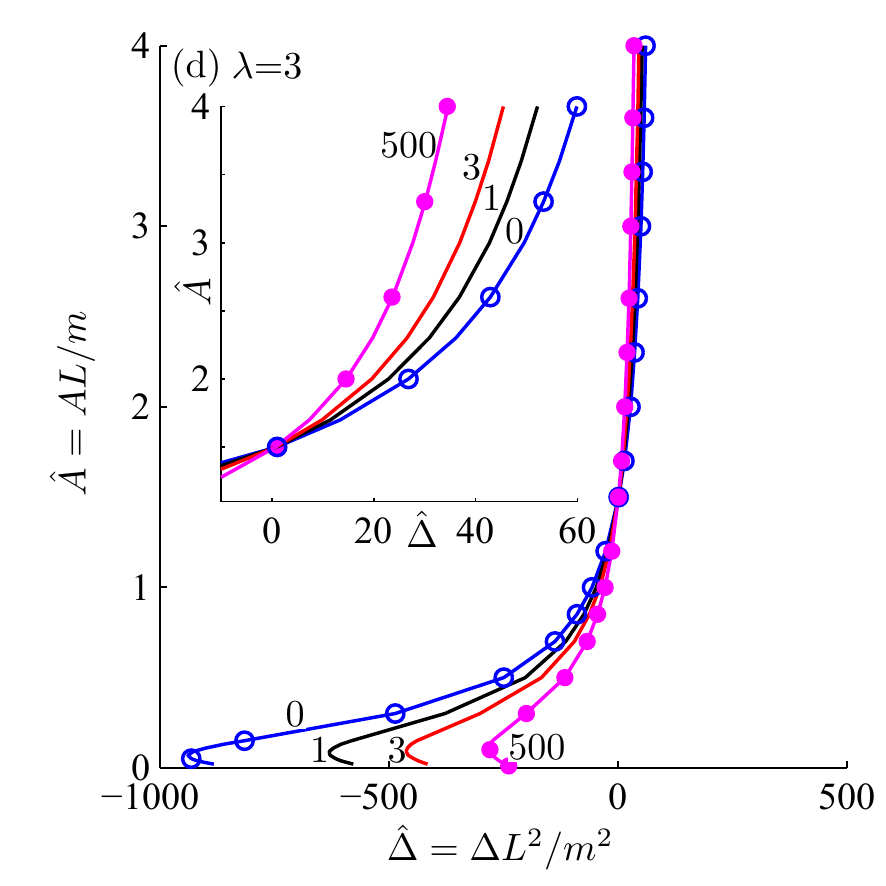}  
\end{multicols}
\caption{Adhesive contact of beams on flexible supports with an adhesive zone model. Top row reports the variation of contact area $\hat{A}$ with total load $\hat{P}$, while the bottom row plots the change of $\hat{A}$ with the punch's displacement $\hat{\Delta}$. Different torsional spring stiffnesses $k_t^f$ are considered,  and they are noted next to their associated curves.  Two different adhesive strengths $\lambda$ are considered, as indicated. The beam's thickness $h=4$ mm and half-span $l=40$ mm. The insets in (b) and (d) depict behavior at high $\hat{A}$. Open and filled circles represent results for a simply supported and a clamped beam, respectively, at the corresponding $\lambda$.}
\label{flexiblebeam_Maugis_diff_ktf_plots}
\end{center}
\end{figure}

From previous discussions, it is expected that results for beams with flexible supports will lie between those obtained for clamped and simply supported beams. Hence, we do not explore this parameter space in great detail. We only consider the variation of $\hat{P}$ and $\hat{\Delta}$ with $\hat{A}$ for several values of torsional stiffness $k_t^f$ for a beam with $h=4$ mm and $l=40$ mm. Two different adhesive strengths $\lambda$ are investigated. The results are shown in Fig.~\ref{flexiblebeam_Maugis_diff_ktf_plots}. The vertical translational spring's stiffness $ k_s^f$ is set to infinity. When $k_t^f=0$, the solutions match with those of a simply supported beam with the corresponding $\lambda$. With increase in $k_t^f$, the solution curves move towards those obtained for a clamped beam and will coincide when $k_t^f$ becomes infinity. It is seen in Fig.~\ref{flexiblebeam_Maugis_diff_ktf_plots} that curves for different $k_t^f$ intersect with each other due to rotation permitted at the supports by the torsional springs -- greater the rotation allowed, higher the displacements, and lower the loads for the same contact area. This intersection point moves up with increasing $\lambda$, as strong adhesive forces are able to bend the beam upwards more easily.

\begin{figure}
\centering
\begin{multicols}{2}
\subfloat[][]{\includegraphics[width=\linewidth]{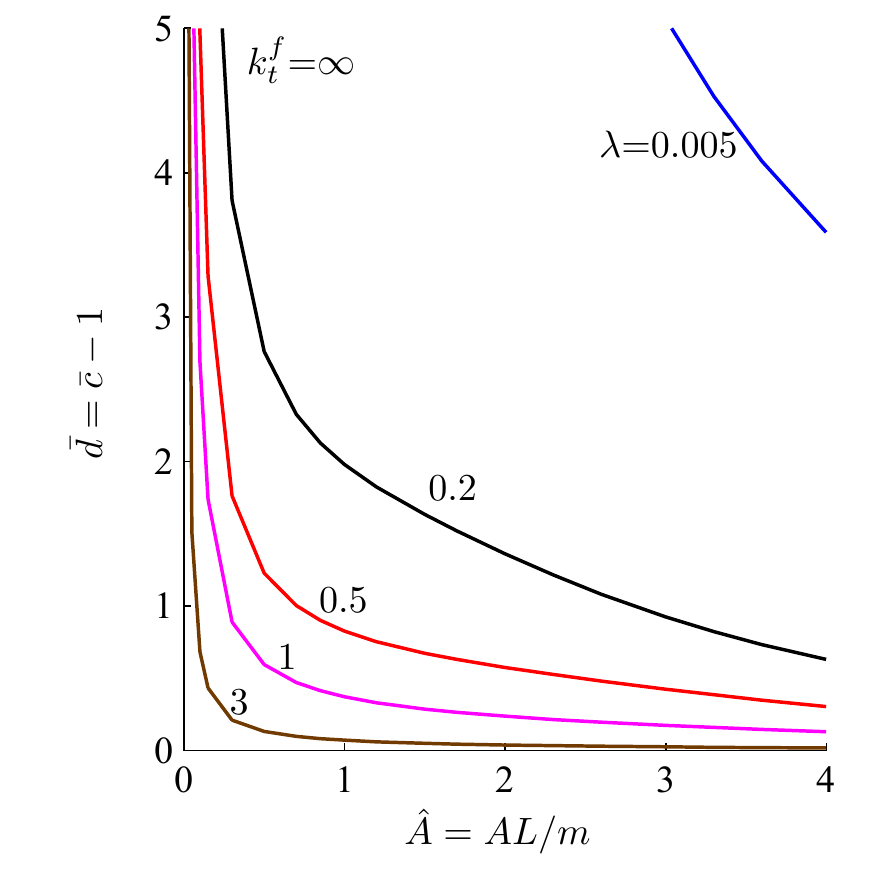}}
\subfloat[][]{\includegraphics[width=\linewidth]{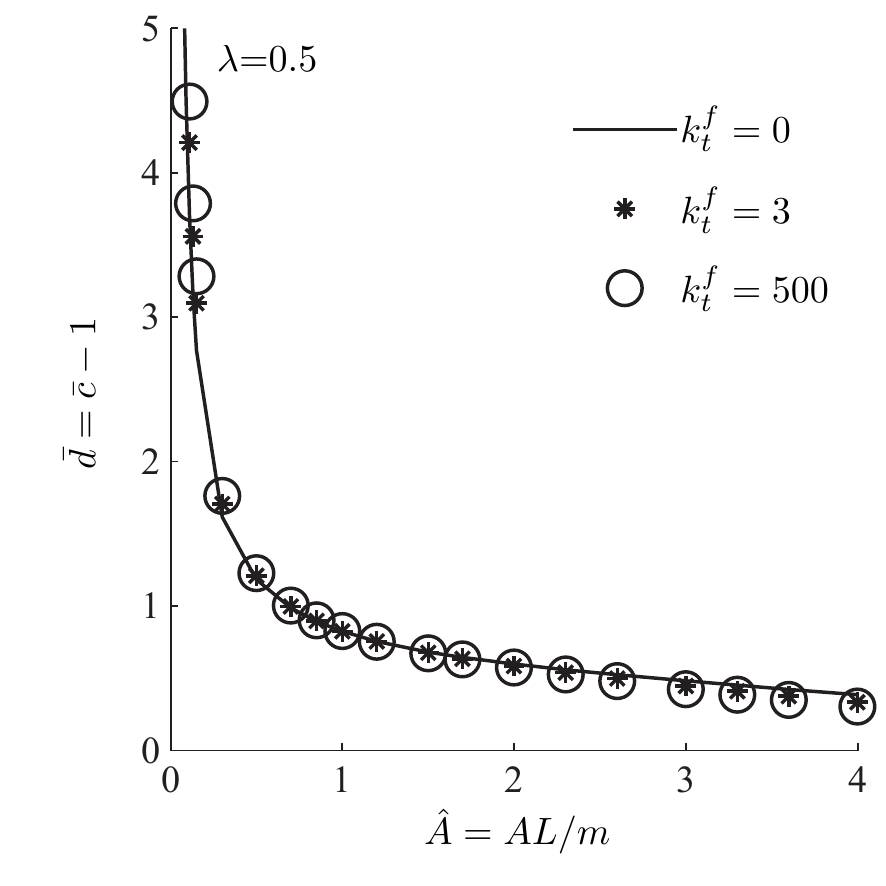}}
\end{multicols}
\caption{Adhesive contact of beams on flexible supports with an adhesive zone model. Variation of the adhesive zone's size $\bar{d}$ with the contact area $\hat{A}$ for (a) different adhesive strengths $\lambda$, with $k_t^f=\infty$ and $k_s^f=\infty$, and (b) three different torsional spring stiffnesses $k_t^f $ at $\lambda=0.5$.}
\label{simplebeam_maugis_h2250_adhesive_zone_size_plot}
\end{figure}

Finally, in Fig.~\ref{simplebeam_maugis_h2250_adhesive_zone_size_plot} we plot the variation of the non-dimensional adhesive zone size $\bar{d} = \bar{c} -1$ with the contact area $\hat{A}$ for different adhesive strengths $\lambda$ and various $k_t^f$. We observe that the adhesive zone size is large for smaller $\lambda$, and decreases with increasing $\lambda$, finally vanishing as $\lambda \rightarrow \infty$. Moreover, we find that the adhesive zone's size does not vary much with the slenderness ratio $l/h$, and spring stiffnesses $k_s^f$ and $k_t^f$.

\section{Experiments and Applications}
In this final section, we present preliminary experimental results on a clamped beam, as well as apply our methods to model structural adhesives of the type shown in Fig.~\ref{fig:struct_adhesive_model}(a). 

\subsection{Experiments}
We have experimentally investigated the indentation of a clamped PDMS (poly-dimethyl-siloxane) beam by a cylindrical glass punch. To make PMDS samples, a uniform mixture of Sylgard 184 silicone elastomer base and curing agent is prepared by taking them in 10:1 weight ratio. Air is desiccated from this mixture, which is then poured into a rectangular mould of desired size. This is cured at room temperature ($\approx 17 ^o$C) for two days and the PDMS sample is extracted from the mould.

\begin{figure}[htbp]
\centering
\subfloat[][]{\includegraphics[width=0.5\linewidth]{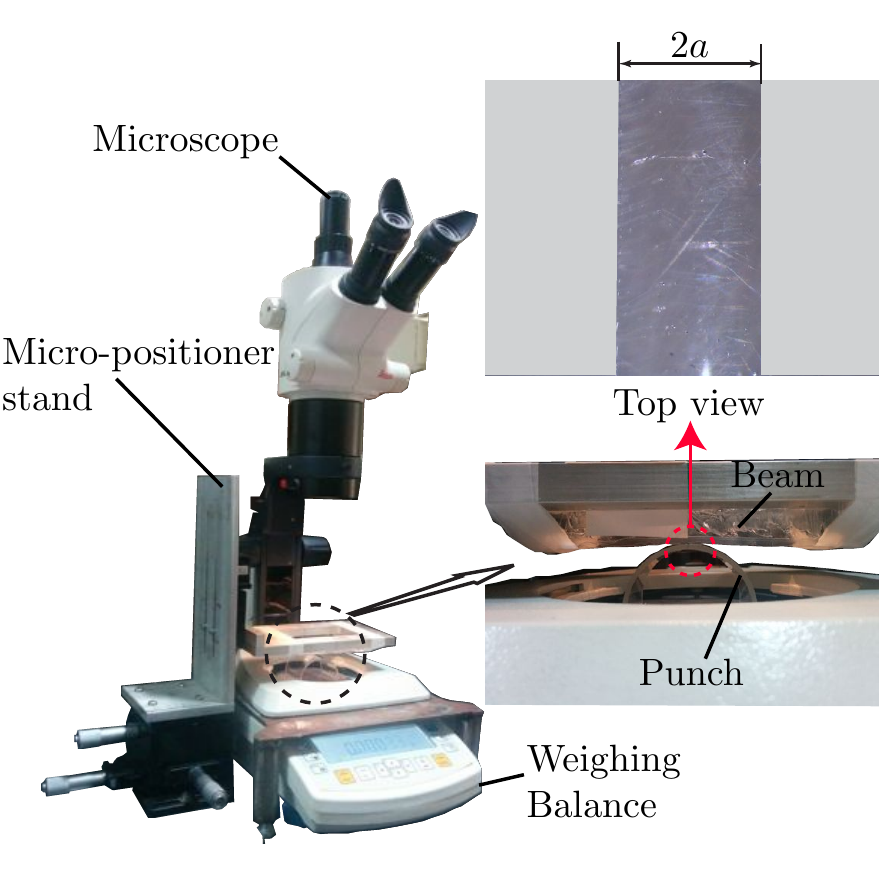}} 
\subfloat[][]{\includegraphics[width=0.5\linewidth]{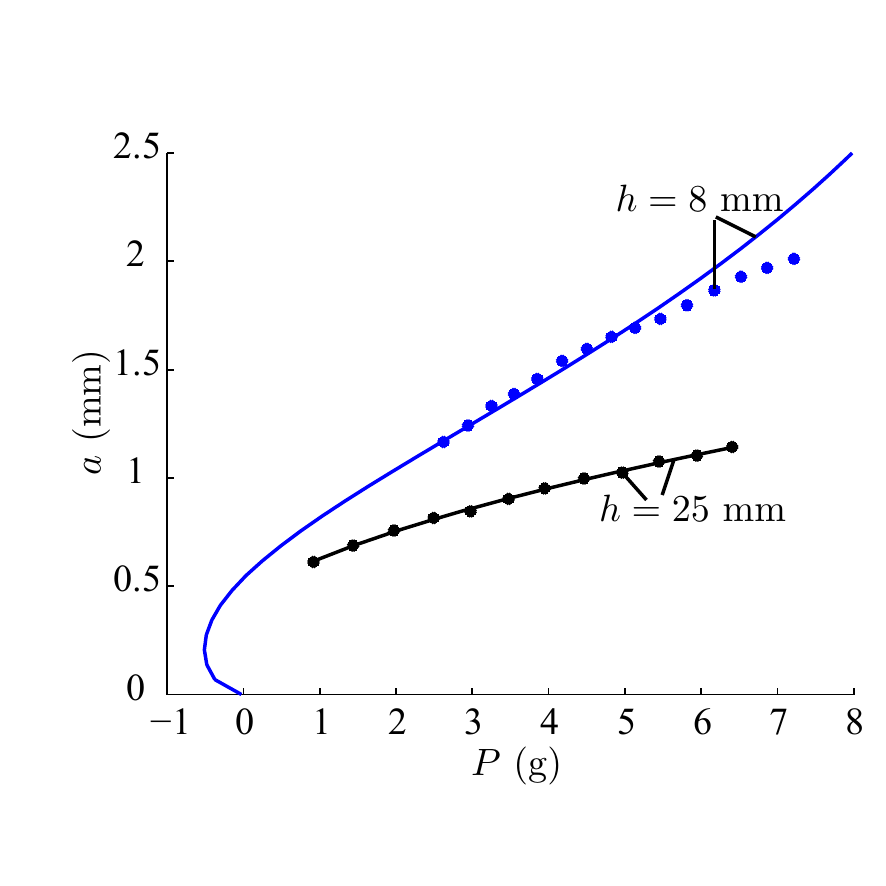}} 
\caption{ (a) Indentation experiments for an adhesive clamped beam of thickness $h=8$ mm and half-span $l=50$ mm. Insets show a closeup of the side view of the indentation and a top view of the contact patch. (b) Variation of the contact area $a$ (in mm) with the total load $P$ (in g). Filled circles represent experimental data. Solid lines correspond to theoretical predictions. For $h=8$ mm, we followed Sec.~\ref{sec:Results_JKR}, while $h=25$ mm, we employed the standard JKR solution for a half-space \citep{chaudhury1996adhesive}.}
\label{fixedbeam_JKr_experiments}
\end{figure}

First, micro-tensile tests are carried out on the PDMS samples to measure their Young's modulus $E$. This is found to be in the range of 1 -- 2 MPa. Next, standard JKR  indentation \citep{chaudhury1996adhesive} tests are carried out. In these tests, PDMS samples of thickness $h \approx 25$ mm are rigidly attached to the micro-positioner, and a cylindrical glass punch of radius $R \approx 27.5$ mm is placed on top of the semi-micro balance; see Fig.~\ref{fixedbeam_JKr_experiments}(a). The PDMS samples are brought into contact with the glass punch employing the micro-positoner. The rectangular contact patch thus formed, shown in the upper inset of Fig.~\ref{fixedbeam_JKr_experiments}(a),  is observed through a microscope, and the load acting on the punch is noted from the semi-micro balance. The Young's modulus $E$ and the work of adhesion $w$ are then found following \cite{chaudhury1996adhesive}.  The JKR experiments confirm the range for $E$ found from micro-tensile tests, and find the work of adhesion $w \approx 27$ mJ/mm$^2$. The values of $E$ and $w$ compare well with those reported earlier by \cite{johnston2014mechanical}, and \cite{Arul2008bioinspired}.

Finally, contact mechanics experiments are carried out on a clamped PDMS beam. In these experiments, the beam's half-span and thickness are maintained at $l \approx 50$ mm and $h \approx 8$ mm, respectively. The total load $P$ and contact patch width $2a$ are measured. The results obtained are then compared with the predictions of our semi-analytical approach in Fig.~\ref{fixedbeam_JKr_experiments}(b). We find good agreement. We observe that the results for a beam vary considerably from that of a half-space lending support for the necessity of the theoretical development presented in this paper. 

More thorough experiments, where we vary parameters like flexibility of end supports, beam thickness $h$, the work of adhesion $w$, etc. are under progress. 

\subsection{Application}
\begin{figure}[htbp]
\centering
\subfloat[][] {\includegraphics[width=0.5\linewidth]{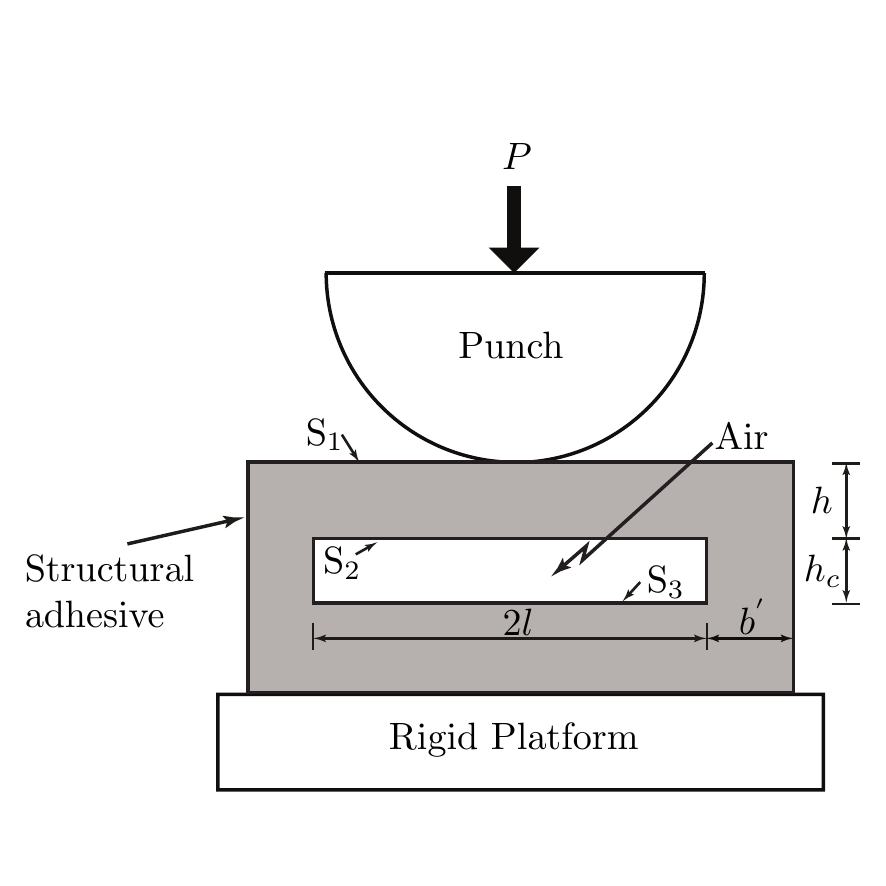}}
\subfloat[][] {\includegraphics[width=0.5\linewidth]{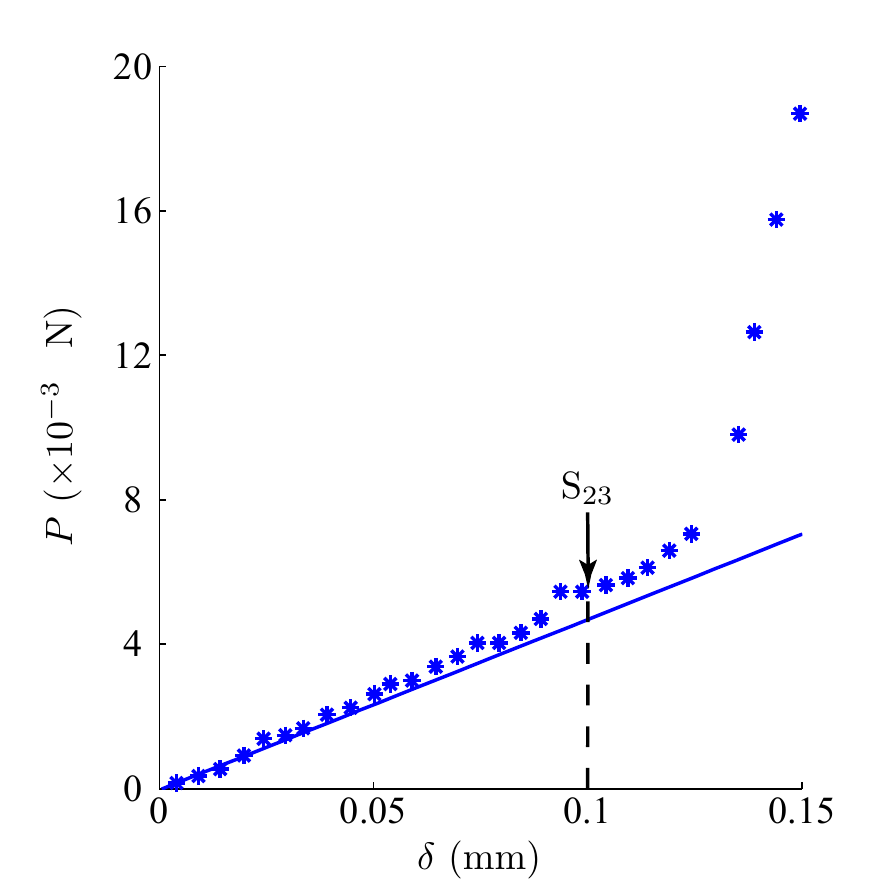}} 
\caption{(a) A structural adhesive with one micro-channel. (b) Variation of the total load $P$ with the punch's displacement $\delta$. The solid line represents the solution obtained from the procedure of Sec.~\ref{sec:Results_JKR}. Asterisk (*) are the experiemtal results of \cite{Arul2008bioinspired}.}
\label{fixedbeam_ghatak}
\end{figure}

We now demonstrate the application of our semi-analytical procedure to the indentation of structural adhesives with one micro-channel, as shown in Fig.~\ref{fixedbeam_ghatak}(a). To this end, the parameters shown in Table~\ref{Ghatak_parameters} are employed to generate our theoretical results. The stiffnesses of the flexible end supports are estimated from a strength-of-materials approach to be $k_s^f \approx 12 b^{'}  l^{3} / h_c h^3$ and $k_t^f \approx l {b^{'}}^{3}/h_c h^3$, where the various geometrical parameters are indicated in Fig.~\ref{fixedbeam_ghatak}(a). Here we have assumed $b^{'} > h$, as $b^{'}$ is \textit{not} reported by \cite{Arul2008bioinspired}.  Our results are then compared with the experimental results of \cite{Arul2008bioinspired} in Fig.~\ref{fixedbeam_ghatak}(b). We find good agreement upto an indentation depth $\delta \approx 0.1$ mm, i.e. until the point S$_{23}$. At this point, the bottom surface S$_2$ begins to interact with the surface S$_3$ in experiments. This feature is not yet implemented in our mathematical model, so that it is expected that our predictions will deviate from experimental observations.

\begin{table}
\begin{center}
\begin{tabular}{ |l|l| } 
 \hline
Geometrical and material parameters & Value \\
\hline 
Beam thickness & $h=0.8$ mm \\
Micro-channel's thickness & $h_c=0.1$ mm \\
Beam's length & $2l=5-8$ mm \\
Punch radius &  $R=2.24$ mm \\
Punch length & $l_p=2.7$ mm \\ 
Shear modulus & $G=1$ MPa \\
Poission's ratio & $\nu=0.49$ \\
Young's modulus & $E=2 \left( 1 + \nu \right) \approx 3$ MPa \\
Work of adhesion & $w= 0.045 \times 10^{-3} $ mJ/mm$^2$ or N/mm \\
\hline
\end{tabular}
\end{center}
\caption{Geometrical and material parameters considered for modeling adhesives with one micro-channel; see also Fig.~\ref{fixedbeam_ghatak}(a). These values are taken from \cite{Arul2008bioinspired}. }
\label{Ghatak_parameters}
\end{table}

\section{Conclusions}
In this article we present a theoretical model for the indentation of adhesive beams mounted on flexible supports. Adhesion, when present, was incorporated through either the JKR approximation or an adhesive zone model. We approximate the displacement of the bottom surface of the beam through Euler-Benoulli beam theory. This is then utilized to formulate a boundary value problem, which is reduced to a single Fredholm integral equation of the first kind for the unknown contact pressure. The integral equation is then solved through a Galerkin projection employing Chebyshev polynomials. Finite element (FE) simulations were carried out for clamped and simply supported non-adhesive beams, and our results compared well with FE predictions, as well as with previously reported theoretical results. Results for adhesive contact were found for several combinations of adhesive strengths, beam geometries, and support flexibilities characterized through torsional and vertical translational springs. Theoretical results for adhesive clamped beam were compared with preliminary experiments and a satisfactory match was observed. Finally, we demonstrated the application of our approach to model a complex structural adhesive.  

The semi-analytic technique presented here assumes that the displacement $v_b$ of the beam's bottom surface is approximated well by Euler--Bernoulli beam theory. This assumption is seen to hold for contact areas less than or equal to the beam's thickness, i.e. $a/h \lesssim 1$. For deeper indentations, we need to formulate the contact problem in terms of two unknowns, viz. the contact pressure $p(x)$ and the displacement $v_b(x)$, and then solve the ensuing dual integral equations. This is done in the second part of this work. We will report extensive experiments on the adhesive contact of beams in the third part of this series. Looking further forward, we envisage extending our technique to modeling adhesive interaction of one beam with another, as is observed in multi-layered structural adhesives shown in Fig.~\ref{fig:struct_adhesive_model}. The present framework may also be adapted to three-dimensional axi-symmetric adhesive contact of punches with plates.

\section{Acknowledgements}
We thank Dr. T. Bhuvana, Prof. P. Venkitanarayanan and  Prof. S. L. Das of the Department of Mechanical Engineering, IIT Kanpur, for help with experiments. We acknowledge partial financial support from the Department of Mechanical Engineering, IIT Kanpur. We are also grateful to Prof. Akash Anand from the Department of Mathematics and Statistics, IIT Kanpur for useful discussions.

\appendix
\section{Calculations using beam theory}
\label{sec:Appendix_beamtheory}
\begin{figure}[htbp]
\centering
\includegraphics[scale=1]{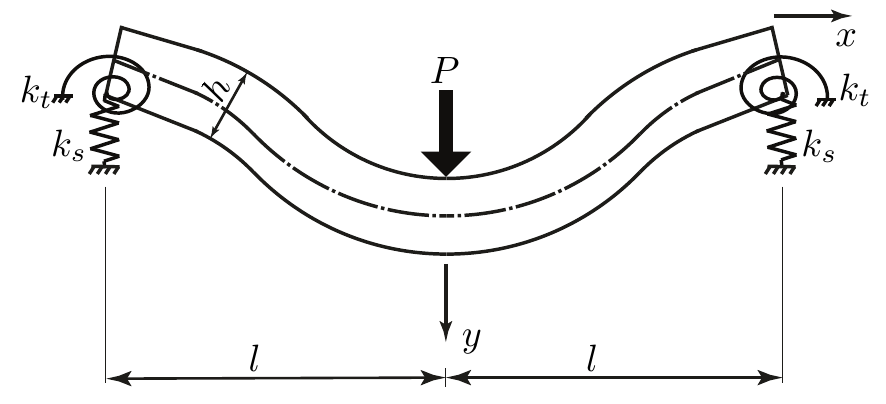}
\caption{An Euler-Bernoulli beam on flexible supports acted upon at its center by a concentrated force $P$.}
\label{fig:fixedbeam_ptload_disp}
\end{figure}

In this section, we find the displacement of an Euler-Bernouli beam subjected to a point load at its center and resting on flexible supports, as shown in Fig.~\ref{fig:fixedbeam_ptload_disp}.

From Euler-Bernoulli beam theory \citep[p.~165, 523, 543]{Crandall2008mech} we find
\begin{equation}
\label{beam_ode}
EI \,  \frac{d^4v_n \left( x \right)}{dx^4} = P \langle x \rangle ^{-1},
\end{equation} 
where $v_n \left( x \right)$ is the displacement of the neutral axis (dash-dot line in Fig.~\ref{fig:fixedbeam_ptload_disp}) and
\begin{equation}
 \langle x \rangle = \left\{ \begin{array}{ll}
 x,  & \quad x >0 \\
 0,  & \quad x  \le 0.
 \end{array} \right. \nonumber
\end{equation}
The appropriate boundary conditions are: 
\begin{subequations}
\label{flexible_end_conditions}
\begin{alignat}{3}
\left. v_n \right|_{x=-l} &=  \frac{P}{2 k_s}, &\quad \left. v_n \right|_{x= l} &= \frac{P}{2 k_s},\\
\left. k_t \frac{dv_n}{dx} \right|_{x=-l} &= \left. E I \frac{d^2v_n}{dx^2} \right|_{x=-l}  &\quad \text{and} \quad \left. -  k_t \frac{dv_b}{dx} \right|_{x=l} &=  \left. E I \frac{d^2v_n}{dx^2} \right|_{x=l}.
\end{alignat}
\end{subequations}

Identifying the displacement $v_b$ of the beam's bottom surface with $v_n$, as is done in beam theory, and solving \eqref{beam_ode} and \eqref{flexible_end_conditions}, provides
\begin{align}
\label{flexible_beam_disp}
v_b \left( x \right) =& \frac{P}{EI} \, \left\{ \frac{1}{6} \langle x \rangle ^{3} - \frac{1}{12} x^3  - \frac{x^2 l}{8} \left( 1 + \frac{E I}{k_t l + E I} \right)  \nonumber \right. \\
& \left. + \frac{1}{24} \left( l^3 + 3 l^3  \frac{E I}{k_t l + E I} + 12 \frac{E I}{k_s} \right) \right\}.
\end{align}
As the displacement is symmetric in $x$, we henceforth employ $v_b \left( x \right)$ for $x \ge 0$ in our calculations.

Non-dimensionalizing \eqref{flexible_beam_disp} following Sec.~\ref{sec:Non-dimensionalization}, we obtain
\begin{equation}
\label{flexiblebeam_disp_nondim}
\vartheta_b \left(  \hat{\tau} \right) = \frac{4 \bar{P} }{3 \bar{I} \left( 1-\nu^2 \right)} \, \, \left\{ \frac{\langle \hat{\tau} \rangle ^{3}}{6}   - \frac{\hat{\tau}^3}{12}   - \frac{\hat{\tau}^2}{8} \left(1+ K_t^{-1} \right)  + \frac{1}{24} \left( 1 + 3  K_t^{-1} + 12  K_s^{-1} \right) \right\},
\end{equation}
where $K_t^{-1} = {E I}/ ({k_t l + E I}) = (1 + k_t^f)^{-1}$ and $K_s^{-1} = {E I}/{k_s l^3} = {k_s^f}^{-1}$. 

When we extended the beam to infinity -- cf. Fig.~\ref{fig:struct_adhesive_model_text2}(b) -- the displacement of the overhang, i.e. $ \hat{\tau} > 1$, is given by
\begin{equation}
\label{flexiblebeam_overhang_formula}
\vartheta_b \left( \hat{\tau} \right)  = \left. \vartheta_b \left( \hat{\tau}  \right) \right|_{\hat{\tau}=1}  + \left. \frac{d \vartheta_b \left( \hat{\tau} \right)}{d \hat{\tau} } \right|_{\hat{\tau}=1} \, \left(  \hat{\tau} - 1 \right).
\end{equation}
Calculating the displacement and slope at $ \hat{\tau} = 1$ from \eqref{flexiblebeam_disp_nondim} and substituting  in \eqref{flexiblebeam_overhang_formula} yields
\begin{equation}
\label{flexiblebeam_overhang_disp}
\vartheta_b \left( \hat{\tau} \right) =  \frac{4 \bar{P} }{3 \bar{I} \left( 1-\nu^2 \right)} \,\left[  \frac{1}{2K_s} - \frac{1}{4K_t} \left(  \hat{\tau} - 1 \right)  \right],  \quad  \hat{\tau} > 1.
\end{equation}
Finally, the displacement of the extended beam's bottom surface over its entire length may be written as
\begin{equation}
\vartheta_b \left( \hat{\tau} \right)= \frac{4 \bar{P} }{3 \bar{I} \left( 1-\nu^2 \right)} \, \, \vartheta_p \left( \hat{\tau} \right),
\end{equation}
where
\begin{equation}
\vartheta_p \left( \hat{\tau} \right) = \left\{
\begin{array}{l}
\left\{ {\hat{\tau}^3 }/{12}   -{\hat{\tau}^2} \left(1+ K_t^{-1} \right)/{8}  +  \left( 1 + 3  K_t^{-1} + 12  K_s^{-1} \right)/24 \right\},  \quad 0 \leq \hat{\tau} \leq 1 \\
\left\{  {K_s^{-1}}/{2} - {K_t^{-1}} \left(  \hat{\tau} - 1 \right) /4 \right\},  \quad \hat{\tau} > 1.
\end{array}
\right.
\end{equation}


Finally, we evaluate the Fourier transform
\begin{equation}
\label{Defn_vw}
\hat{\vartheta}_b \left( \hat{\omega} \right)  = \int\limits_{-\infty}^{\infty} \vartheta_b \left( \hat{\tau} \right) \cos \left( \hat{\omega} \hat{\tau} \right) \, d \hat{\tau} = 2 \int\limits_{0}^{\infty} \vartheta_b \left( \hat{\tau} \right) \cos \left( \hat{\omega} \hat{\tau} \right) \, d \hat{\tau},
\end{equation}
which is required in \eqref{Int_eqn_final}. The above integrals are typically undefined, as $\vartheta \left(  \hat{\tau} \right)$ is unbounded once the beam is extended to infinity, unless the beam is clamped. To overcome this, we invoke St. Venant's principle by which, displacement of the overhang, sufficiently far away from the supports, may be modified without exerting any significant influence on the displacement and stresses in the portion of the beam lying within the supports. To this end, we modify the displacement of the beam's bottom surface by introducing
\begin{equation}
\label{disp_beam_mollified}
{\vartheta}^{M}_b \left( \hat{\tau} \right) = \vartheta_b \left( \hat{\tau} \right)  \, \, \cdot \, \, W \left( \hat{\tau} \right),
\end{equation}
with
\begin{equation}
\label{defn:mollifiers}
W \left( \hat{\tau} \right) = \left\{ 
\begin{array}{ll}
1, &  \text{for } \hat{\tau} \leq  \hat{l}_1, \\
 w_2 \left( \hat{\tau} \right),  & \text{for }\hat{l}_1<\hat{\tau}< \hat{l}_2 \\
0, & \text{for } \hat{\tau} \geq \hat{l}_2,
\end{array}
,\right.
\end{equation}
where
\begin{equation}
\label{defn:mollifiers2}
w_2 \left( \hat{\tau} \right) = 
\frac{ \exp \left\{ {-1/\left( \hat{l}_2 - \hat{\tau} \right)^2} \right\} }{\exp \left\{ {-1/\left( \hat{l}_2 - \hat{\tau} \right)^2} \right\} + \exp \left\{ {-1/\left( \hat{\tau} - \hat{l}_1 \right)^2} \right\} },
\end{equation}
and $\hat{l}_1$ and $\hat{l}_2$ locate points on the beam that are far away from its supports, i.e. $\hat{l}_2>\hat{l}_1>>1$. The function $W \left( \hat{\tau} \right)$ is a \emph{mollifier}, see \cite{Muthukumar}, and is infinitely differentiable everywhere.

The mollified displacement ${\vartheta}^{M}_b \left( \hat{\tau} \right)$  in \eqref{disp_beam_mollified} is now utilized in \eqref{Defn_vw} to compute the Fourier transforms. Thus,
\begin{equation}
\label{four_defn_v}
\hat{\vartheta}_b \left( \hat{\omega} \right)  \approx 2 \int\limits_{0}^{\infty} {\vartheta}^M_b \left( \hat{\tau} \right) \cos \left( \hat{\omega} \hat{\tau} \right) \, d \hat{\tau} = 2 \int\limits_{0}^{\infty} \vartheta_b \left( \hat{\tau} \right) W \left( \hat{\tau} \right) \cos \left( \hat{\omega} \hat{\tau} \right) \, d \hat{\tau}.
\end{equation}
Evaluating the above integral and replacing the total load $\bar{P}$ from \eqref{total_load_cheby} in the resulting equation provides
\begin{align}
 \hat{\vartheta}_b \left( \hat{\omega} \right)   =  \left\{ \frac{4 \pi b_0 }{3  \bar{I} \left( 1-\nu^2 \right)}  - \frac{4 \lambda A m \bar{c} }{3 \hat{\gamma}^3 \bar{I} L \left( 1-\nu^2 \right)} \right\} \hat{\vartheta}_p \left( \hat{\omega} \right),
 \label{fourier_v}
\end{align}
where
\begin{equation}
\label{four_defn_vp}
\hat{\vartheta}_p \left( \hat{\omega} \right)  = 2 \int\limits_{0}^{\infty} \vartheta_p \left( \hat{\tau} \right) W \left( \hat{\tau} \right) \cos \left( \hat{\omega} \hat{\tau} \right) \, d \hat{\tau}.
\end{equation}
Setting $k_t^f \rightarrow \infty$ and $k_s^f \rightarrow \infty$, we obtain results for a clamped beam, while those for a simply supported beam are found by taking $k_t^f \rightarrow 0$ and $k_s^f \rightarrow \infty$.

Finally, we write
\begin{align}
\frac{1}{\pi \hat{\gamma}} \hat{\vartheta}_b \left( \hat{\omega} \right)  =  \left\{ \frac{4 b_0 }{3 \bar{I} \left( 1-\nu^2 \right) \hat{\gamma} }  - \frac{4 \lambda A m \bar{c} }{3 \pi \hat{\gamma}^4 \bar{I} L \left( 1-\nu^2 \right)} \right\} \hat{\vartheta}_p \left( \hat{\omega}  \right). \label{V_omega_defn}
\end{align}

\section{Vertical displacement of the beam's top surface in Fourier space $V \left( \xi, 0 \right)$ }
\label{sec:top_surface_disp}
In Fourier space, the transformed horizontal and vertical displacements may be solved as, respectively, 
\begin{align}
\label{Gen_disp_1}
U \left( \xi, y \right) &= \left\{ \kappa a_1 + \xi y \left( a_1+ i a_3 \right) \right\} e^{\xi y}+\left\{ \kappa b_1 - \xi y \left( b_1 - i b_3 \right) \right\} e^{-\xi y}, \\  
\text{and } \quad
\label{Gen_disp_2}
V \left( \xi, y \right) &= \left\{ \kappa a_3 + i \xi y \left( a_1+ i a_3 \right) \right\} e^{\xi y}+\left\{ \kappa b_3 + i \xi y \left( b_1 - i b_3 \right) \right\} e^{-\xi y},
\end{align}
with
\begin{equation}
U \left( \xi, y \right) = \int\limits_{-\infty}^{\infty} u \left( x, y\right) e^{i \xi x} dx \quad \text{and} \quad
V \left( \xi, y \right) = \int\limits_{-\infty}^{\infty} v \left( x, y\right) e^{i \xi x} dx, \nonumber
\end{equation}
where $a_1$, $a_3$, $b_1$ and $b_3$ are unknown constants. These constants are obtained by satisfying boundary conditions \eqref{BC_for_contact_x},  which in Fourier space are 
\begin{subequations}
\label{BCS_for_contact_xi}
\begin{alignat}{3}
\text{at} \; \;  y=0: &\quad S_{\xi y} &= 0,  &\quad S_{yy} &= \bar{P}_c (\xi), \label{BC_y0_xi}\\ 
\text{and at} \; \;  y=h: &\quad S_{\xi y} &= 0,  &\quad V &= \bar{v}_b (\xi), \label{BC_yh_xi}
\end{alignat}
\end{subequations}
with
\begin{eqnarray}
S_{\xi y} &=& \frac{E}{2 \left( 1+ \nu \right)} \left( \frac{d}{dy} U - i \xi V\right),  \nonumber\\
S_{yy} &=& \frac{E}{ \left( 1+ \nu \right)} \left\{ \frac{d}{dy} V + \frac{\nu}{1-2 \nu} \left(-i \xi U + \frac{d}{dy} V \right) \right\},  \nonumber \\
\bar{P}_c (\xi) &=& \int\limits_{-\infty}^{\infty} -P_c \left( x \right) e^{i \xi x} dx  \quad \text{and } \quad 
\bar{v}_b (\xi) = \int\limits_{-\infty}^{\infty} v_b \left( x \right) e^{i \xi x} dx. \nonumber 
\end{eqnarray}
Solving \eqref{Gen_disp_1}--\eqref{BCS_for_contact_xi}, we obtain the vertical displacement of the beam's top surface in Fourier space as
\begin{align}
V \left( \xi, 0 \right)=&  - \frac{2 \, \bar{P}_c \left( \xi \right)}{E^*} \, \frac{ \sinh^2{\xi \, h}}{\xi \left( \xi \, h + \sinh{\xi \, h } \cosh{\xi \, h} \right)} + \bar{v}_b \left( \xi \right) \, \frac {\sinh{ \xi \, h} + \xi\,h \cosh{ \xi\,h } }{ \xi \, h + \sinh{ \xi \, h} \cosh{ \xi \, h }}. \nonumber
\end{align}

\section{Evaluation of the integrals $\alpha_{n} \left( \bar{\omega} \right)$}
\label{sec:Appendix_alphan}
We recall from \eqref{alpha_cheby_2} in Sec.~\ref{sec:Numerical_solution} that
\begin{equation}
\label{alpha_n_revisited}
\alpha_{n} \left( \bar{\omega} \right) = \int\limits_{-1}^{1} \frac{1}{\sqrt{ \left( 1-\bar{\tau}^2 \right)}} T_{n} \left( \bar{\tau} \right) \cos \left( \bar{\omega} \tau \right)  \: \: d\bar{\tau}.
\end{equation}
We now compute these integrals explicitly. First, consider odd $n$. For this, the integrand is an odd function so that
\begin{equation}
\alpha_{2n-1} \left( \bar{\omega} \right) = 0.
\end{equation}
Next, evaluating \eqref{alpha_n_revisited} for even $n$ we obtain the first few $\alpha_n$ as
\begin{eqnarray}
\alpha_{0} \left( \bar{\omega} \right) &=& \pi \,{{\rm J}\left(0,\,\bar{\omega}  \right)}, \nonumber\\
\alpha_{2} \left( \bar{\omega}  \right) &=& \pi \,{{\rm J}\left(0,\,\bar{\omega}  \right)} - {\frac {2 \, \pi \, {{\rm J}\left(1,\, \bar{\omega}  \right)}}{\bar{\omega}}} \nonumber, \\
\alpha_{4} \left( \bar{\omega}  \right) &=& \pi \,{{\rm J}\left(0,\,\bar{\omega} \right)}- {\frac {8\, \pi \,{{\rm J}\left(1,\,\bar{\omega} \right)}}{\bar{\omega}}}  - {\frac {24\, \pi \,{{\rm J}\left(0,\, \bar{\omega} \right)}}{\bar{\omega}^{2}}} + {\frac {48\, \pi \, {{\rm J}\left(1,\,\bar{\omega} \right)}}{\bar{\omega}^{3}}}, \nonumber \\
\text{and} \quad
\alpha_{6} \left( \bar{\omega} \right) &=& \pi \,{{\rm J}\left(0,\,\bar{\omega} \right)}- {\frac {18\, \pi \,{{\rm J}\left(1,\,\bar{\omega}\right)}}{\bar{\omega}}}  - {\frac {144\, \pi \, {{\rm J}\left(0,\bar{\omega} \right)}}{\bar{\omega}^{2}}} + {\frac {768\, \pi \, {{\rm J}\left(1,\, \bar{\omega} \right)}}{\bar{\omega}^{3}}} \nonumber \\ &&  + {\frac {1920\, \pi \,{{\rm J}\left(0,\,\bar{\omega}\right)}}{\bar{\omega}^{4}}}  - {\frac {3840\, \pi \,{{\rm J}\left(1,\,\bar{\omega} \right)}}{\bar{\omega}^{5}}}, \label{alpha_n_calcs}
\end{eqnarray}
where ${{\rm J}\left(n,\, \bar{\omega} \right)}$ are the Bessel's functions of the first kind of order $n$. Employing the recurrence relation (\citealt[p.~1016]{polyanin2008handbook}),
\begin{equation}
{{\rm J}\left(n+1,\, \bar{\omega} \right)} = \frac{2n}{\bar{\omega}} {{\rm J}\left(n,\, \bar{\omega} \right)} - {{\rm J}\left(n-1,\,\bar{\omega} \right)}, \nonumber
\end{equation}
we rewrite \eqref{alpha_n_calcs} as
\begin{equation}
\alpha_{2} \left( \bar{\omega} \right) = - \pi \,{{\rm J}\left(2,\, \bar{\omega} \right)}, \quad
\alpha_{4} \left( \bar{\omega} \right) =  \pi \,{{\rm J}\left(4,\, \bar{\omega} \right)} \quad \text{and} \quad
\alpha_{6} \left( \bar{\omega} \right) = - \pi \,{{\rm J}\left(6,\, \bar{\omega} \right)}. \nonumber
\end{equation}
In general, it is possible to show that
\begin{align}
\alpha_{2n} \left( \bar{\omega} \right) &= \left( -1 \right)^n \pi \,{{\rm J}\left(2n,\, \bar{\omega} \right)}. 
\end{align}

\singlespacing
\providecommand{\noopsort}[1]{}\providecommand{\singleletter}[1]{#1}%


\end{document}